\documentclass[aps,prb,showpacs,preprint,preprintnumbers,eqsecnum]{revtex4}
\usepackage{amssymb}
\usepackage{amsfonts}
\usepackage{amsmath}
\usepackage{graphicx}
\usepackage{dcolumn}
\usepackage{bm}

\input{tcilatex}

\begin{document}

\preprint{\textbf{2008-11-13}}
\title{ Spin rotation, spin filtering, and spin transfer in directional
tunneling through non-centrosymmetric semiconductor barriers}
\author{T.\ L.\ Hoai Nguyen$^{a,b}$}
\author{Henri-Jean Drouhin$^{a}$}
\email{Henri-Jean.Drouhin@polytechnique.edu}
\author{Jean-Eric Wegrowe$^{a}$}
\author{Guy Fishman$^{b}$}
\affiliation{$^{a}$ Ecole Polytechnique, LSI, CNRS and CEA/DSM/DRECAM, Palaiseau, ~F-91128}
\affiliation{$^{b}$Univ Paris-Sud, IEF, CNRS,~Orsay, F-91405}
\date{\today }

\begin{abstract}
We discuss possible tunneling phenomena associated with complex wave vectors
along directions where the spin degeneracy is lifted in non-centrosymetric
semiconductors. We show that the result drastically depends on the
direction. In the [110] direction, no solution can be calculated in the
usual way assuming that the wave function and its derivative are continuous.
Conditions that permit one to find physical solutions are discussed and
consequences are drawn. As a result, there is no spin filtering in such a
direction but the spin undergoes a precession through the barrier with the
rotation angle being proportional to the barrier thickness. In a direction
close to [001] we find a spin-filter effect in close agreement with the
model discussed by Perel et al. [Phys. Rev. B \textbf{67}, 201304(R)
(2003).].
\end{abstract}

\maketitle

\section{Introduction}

\label{Intro}

Understanding spin-dependent tunneling through semiconductor barriers is a
fundamental problem in semiconductor physics. A description of this coherent
process is crucial for spin-subband engineering of semiconductor
heterostructures and superlattices. Moreover, spin-dependent tunneling
through crystalline barriers has also become a topic of major interest in
spintronics.\cite{Butler,Elsen} This article thus lies at the interface
between general semiconductor physics and spintronics. Regarding
semiconductor physics, it has close connections with open problems in the
envelope function theory.\cite{LK} Harrison studied the problem of
heterogeneous materials and introduced the conditions of discontinuity of
the envelope function, taking a general viewpoint, well beyond the
semiconductor area.\cite{Harrison} A decisive step in semiconductors was
performed by BenDaniel and Duke\cite{BDD} who defined specific discontinuity
conditions of the derivative of the envelope function between two media with
different effective masses, based on the conservation of the probability
current. This approach has been successfully applied to heterostructures by
Bastard\cite{Bastard} and has become the standard calculation routine,
yielding very accurate energy positions of the energy bands, in perfect
agreement with the experimental data.\cite{Balian} Hereafter, analogously,
we deal with periodic lattices which are perturbed by a spin-orbit potential
and where, due to the absence of space-inversion symmetry, the spin
degeneracy of the bands is lifted through a wave-vector-dependent
\textquotedblleft exchange\textquotedblright\ field.\cite{DyaPer,Slonczewski}
We show that the matching conditions of the derivative of envelope function
at the boundaries cannot be \textquotedblleft as usual\textquotedblright .
Thus, the basic tunneling equations are not known.

First of all, dealing with tunneling phenomena requires an accurate
knowledge of the energy structure in the forbidden band gap i.e. of the
complex band structure of the barrier material. In pioneering articles, Heine%
\cite{Heine} and Jones\cite{Jones} derived general properties of the
evanescent states and showed their complexity over 6-dimensional wave-vector
space, consisting of complex vectors associating a pure imaginary component
to a real - propagating - one. It is often thought that the electrons will
tunnel through such complex-wave-vector states as they would do through
usual evanescent states (with pure imaginary wave vectors), and this
intuitive explanation is probably supported by our familiarity with the
tunneling of electrons located in semiconductor side valleys (e.g. in the
conduction band of silicon). Hereafter, we deal with spin-dependent
tunneling of conduction electrons through a gallium arsenide barrier, a
compound with no inversion symmetry.\cite{DRE} Such processes were
investigated by Perel' et al. in a stimulating article, using the effective
mass approximation and under simplifying assumptions; Quite large
spin-filter effects were predicted.\cite{Perel03} Although the complex band
structure of GaAs was expected to be well known, we have recently found
that, in fact, the spin-orbit interaction and the absence of inversion
symmetry had never been taken into account simultaneously throughout the
Brillouin zone.\cite{Rougemaille,Richard} The evanescent band structure was
calculated by several authors. Chang\cite{Chang} considered semiconductors
oriented in the [100], [111], and [110] directions, with space-inversion
centers ($O_{h}$ group) or without space-inversion centers ($T_{d}$ group),
but without taking into account the spin-orbit coupling. Chang and Schulman%
\cite{ChangSchulman} performed a detailed calculation of the band structure
of silicon, which belongs to the $O_{h}$ group. Schuurmans and t'Hooft\cite%
{SchuurHooft} studied semiconductors belonging to the $T_{d}$ group but
explicitly discarded terms which lead to odd $k$ terms so that essentially
they studied GaAs and AlAs as if they belonged to the $O_{h}$ group. In Ref. 
%TCIMACRO{\TeXButton{\onlinecite{Richard}}{\onlinecite{Richard}} }%
%BeginExpansion
\onlinecite{Richard}
%EndExpansion
the evanescent band structure in the fundamental gap of GaAs-like III-V
semiconductors, including both the spin-orbit coupling and the lack of
inversion symmetry, was carefully calculated within a $14\times 14$ and a $%
30\times 30$ $\mathbf{k\cdot p}$ Hamiltonian framework. Then it was
demonstrated that the evanescent states in the fundamental gap present an
original topology, with loops connecting opposite spin states at the center
of the Brillouin zone.\cite{Rougemaille} This very structure has strong
consequences for electron tunneling. Here, in order to remove any
unnecessary complexity, we start dealing with electrons with a unique
effective mass $m$, inside and outside the barrier - an approximation used
in numerical applications in Ref. 
%TCIMACRO{\TeXButton{\onlinecite{Perel03}}{\onlinecite{Perel03}}}%
%BeginExpansion
\onlinecite{Perel03}%
%EndExpansion
. The spin splitting in the barrier is described via the D'yakonov-Perel'
(DP) Hamiltonian.\cite{DyaPer} We thus revisit a classic in elementary
quantum mechanics - the tunneling of free electrons through a square
potential barrier - but in a case where the evanescent states in the barrier
are spin split. From general considerations, we derive relevant boundary
conditions which are sensitive on the crystallographic direction. We
demonstrate that the tunneling process can become rather involved: The case
of loop-shaped real-energy lines correspond to wave vectors which have both
an imaginary component, which defines the tunneling direction, and an 
\textit{orthogonal} real component so that one has to deal, so to say, with
a \textquotedblleft classical\textquotedblright\ tunnel effect in the sense
where it is possible to recover almost usual tunneling properties -
analogous to off-normal tunneling of free electrons - but in a subtle way.
In the case of one-dimensional tunneling with a complex (neither real nor
purely imaginary) wave vector, the tunnel effect seems to be
\textquotedblleft anomalous\textquotedblright : A spin precession occurs
around a \textquotedblleft complex magnetic field\textquotedblright . We
show that the derivative of the envelope function, which is the solution of
the Schr\"{o}dinger equation, undergoes discontinuities at the barrier plane
- usually, in semiconductor heterostructures, discontinuities of the
derivative arise \textit{as a consequence of the different effective masses}
in the well and in the barrier material\cite{Bastard} - and we propose a
treatment of heterostructures. After entangling the two spin channels, it is
possible to recover a situation which has strong analogy with standard
tunneling and where the discontinuity of a \textquotedblleft magnetic
current\textquotedblright\ can be viewed as the result of a kinetic-momentum
transfer at the barrier interfaces. The spin-orbit-split barrier exerts a
torque on the electron spin, similar to spin-torque phenomena in
ferromagnetic junctions as predicted by Slonczewski\cite{Slonczewski} and
Berger\cite{Berger}, but in this latter case, as the barrier is constituted
of magnetic material, a spin transfer occurs between the tunneling electrons
and the magnetization.

The layout of this paper is as follows: In Sec. \ref{back}, we give the
background relative to the spin splitting and to the conservation of the
probability current which will be used afterwards. We show how the spin
splitting can lead to complex (not strictly imaginary) wave vectors in the
barrier and we analyze the consequences on the probability current. In Sec.
III, we study a barrier normal to [110], and in Sec. \ref{ortho}, we look in
detail at the case of an incident wave whose direction is almost normal to a
[001] barrier. A summary is given in Sec. \ref{Co}.

\section{Background\label{back}}

\subsection{Symmetry}

Let us present the notations used throughout the present article (see Fig.
1). $\mathbf{e}$ is a unit vector. The direction of the axes, defined by $%
\mathbf{e}_{x}$, $\mathbf{e}_{y}$, $\mathbf{e}_{z}$ with respect to crystal
axes, will be given in each case. $\mathbf{e}_{z}$ is normal to the barrier. 
$\mathbf{e}_{z}=\mathbf{e}_{110}$ in Sec. III and $\mathbf{e}_{z}=\mathbf{e}%
_{001}$ in Sec. \ref{ortho}. We define $\mathbf{k}_{I}\mathbf{=\xi +q,\ k}%
_{II}=\mathbf{\xi +Q+}i\mathbf{K}$ ($\mathbf{\xi }$, $\mathbf{q}$, $\mathbf{Q%
}$, $\mathbf{K}$ are all real vectors),$\mathbf{\ }$and $\mathbf{k}_{III}%
\mathbf{=\xi +q}$. We also introduce the following notations: $\mathbf{\rho =%
}x\mathbf{e}_{x}+y\mathbf{e}_{y}$, $\mathbf{\xi =}\xi _{x}\mathbf{e}_{x}+\xi
_{y}\mathbf{e}_{y}$, $\mathbf{q}=q\mathbf{e}_{z}$, $\mathbf{Q=}Q\mathbf{e}%
_{z}$, $\mathbf{K=}K\mathbf{e}_{z}$, $\mathbf{k}_{I}\mathbf{\cdot r=k}_{III}%
\mathbf{\cdot r=}\xi _{x}x+\xi _{y}y+qz$, $\mathbf{k}_{II}\cdot \mathbf{r=}%
\xi _{x}x+\xi _{y}y+\left( Q+iK\right) z$. Without spin, the wave function
of the incident plane wave and in the barrier should be written as $%
e^{i\left( \mathbf{\xi \cdot \rho +}qz\right) }$ and $e^{i\left[ \mathbf{\xi
\cdot \rho +}\left( Q+iK\right) z\right] }$ respectively.

To describe the structure of the evanescent states, we use the $\mathbf{%
k\cdot p}$ method. In a $n$-band model, the energy dispersion curves result
from the diagonalization of a $(n\times n)$ $\mathbf{k\cdot p}$ Hamiltonian $%
\widehat{H}$, but $\mathbf{k}$ is a complex vector so that $\widehat{H}$ is
no longer Hermitian and the evanescent states are associated only to real
eigenvalues $E$.\cite{Rougemaille,Richard} To find the energy dispersion
curves, we have to solve the secular equation $\det M(\mathbf{k})=\det [%
\widehat{H}-E\widehat{I}]$, where $\widehat{I}$ is the identity. Because the
Hamiltonian is Hermitian when $\mathbf{k}$ is a real vector, we have the
relation $M(\mathbf{k})^{t\ast }=M(\mathbf{k}^{\ast })$. Thus, $\det M(%
\mathbf{k}^{\ast })=\left[ \det M(\mathbf{k})^{t}\right] ^{\ast }=\left[
\det M(\mathbf{k})\right] ^{\ast }$. It follows that $E_{n}(\mathbf{k}%
)=E_{n^{\prime }}(\mathbf{k}^{\ast })$, where the band indices $n$ and $%
n^{\prime }$ may or may not refer to the same band.\cite{Heine,Jones}
Moreover, Kramers conjugates correspond to the same energy, so that the
state associated to $\left( \mathbf{k},\ \left\vert up\right\rangle \right) $
and the state associated to $\left( -\mathbf{k},\,\left\vert
down\right\rangle \right) $ are degenerate.\cite{Messiah,Kit-I} Let us
recall that Kramers-conjugate states are obtained by application of $%
\widehat{K}$, the time reversal operator: $\widehat{K}=-i\sigma _{y}\widehat{%
K}_{0}$ where $\sigma _{y}$ is the relevant Pauli matrix and $\widehat{K}%
_{0} $ is the operation of taking the complex conjugate.\cite{Messiah} Thus
in GaAs, the spin degeneracy is lifted and we expect that the four states $%
\left[ \left( \mathbf{k,}\,\left\vert s\right\rangle \right) \text{%
,\thinspace }\left( \mathbf{k}^{\ast },\,\left\vert s^{\prime }\right\rangle
\right) \text{,\thinspace }\left( -\mathbf{k}^{\ast },\,\left\vert
-s\right\rangle \right) \text{,\thinspace and }\left( -\mathbf{k}%
,\,\left\vert -s^{\prime }\right\rangle \right) \right] $ be degenerate, $%
\left\vert s\right\rangle $ and $\left\vert s^{\prime }\right\rangle $ being
up-spin states in directions which, generally, are not parallel (Fig. 2). We
are going to see a concrete example in Subsec. \ref{energie}, where $%
\left\vert s\right\rangle $ and $\left\vert s^{\prime }\right\rangle $ are
quantized in the same direction, and in Sec. \ref{ortho} where $\left\vert
s\right\rangle $ and $\left\vert s^{\prime }\right\rangle $ are not
quantized in the same direction.

\subsection{Energy levels\label{energie}}

In the introduction, we mentioned that the evanescent band structure is
deeply altered when the lack of inversion symmetry is taken into account
together with the spin-orbit splitting. A particular topology consisting of
loops connecting Kramers-conjugate spin states near the zone center was
shown along directions of the type $K[\xi /K,\,0,\,i]$ when the ratio $\xi
/K=\tan \theta $ is fixed. Such loop structure can be expected to arise as
it is known that a band cannot stop.\cite{Heine} Depending on $\theta $, we
obtain the different pictures shown in Fig. 3. The spin vector along a loop
is defined by the mean value of the Pauli operator $\widehat{\mathbf{\sigma }%
}$. In the small-$k$ limit, we get two opposite spin vectors. When going off
the zone center, a numerical calculation shows that the two spin vectors
rotate to become parallel at\ the point where the two subbands are
connecting. The appearance of these loops is the fingerprint of a strong
band mixing of the first conduction band and of the three upper valence
bands with remote bands (more precisely with the second conduction band).
Indeed, as long as the wave vector remains in some vicinity of $\Gamma $,
the energy levels are well described by the DP Hamiltonian, where the spin
states in the subbands only depend on $\theta $ (see Sec. \ref{ortho}).
Observe, in Fig. 3, that - because the extension of the loop tends to zero
when $\theta $ tends to $45%
%TCIMACRO{\U{b0}}%
%BeginExpansion
{{}^\circ}%
%EndExpansion
$ - the portion of the loops which can be described in this analytical model
also has an extension which can become vanishingly small. Hereafter, we stay
in the framework of the DP model, which allows analytical calculations.

Throughout the present paper, we take the origin of the energy at the bottom
of the conduction band so that the relevant Hamiltonian is written as%
\begin{eqnarray}
\widehat{H} &=&\widehat{H}_{0}+\widehat{H}_{DP}  \notag \\
\widehat{H}_{0} &=&\frac{\widehat{\mathbf{p}}^{2}}{2m}=\frac{-\hbar ^{2}}{2m}%
\mathbf{\nabla }^{2}=-\gamma _{c}\mathbf{\nabla }^{2}  \label{H0_HDP} \\
\widehat{H}_{DP} &=&\gamma \,\mathbf{\chi }\cdot \widehat{\mathbf{\sigma }} 
\notag
\end{eqnarray}

where $m$ is the effective mass. $\widehat{H}_{DP}$ is the DP Hamiltonian
which describes the $k^{3}$ spin-splitting:\cite{DyaPer} $\mathbf{\chi =\chi 
}\left( \mathbf{k}\right) =\left[ \chi _{x},\,\chi _{y},\,\chi _{z}\right] =%
\left[ k_{x}\left( k_{y}^{2}-k_{z}^{2}\right) ,\,k_{y}\left(
k_{z}^{2}-k_{x}^{2}\right) ,\,k_{z}\left( k_{x}^{2}-k_{y}^{2}\right) \right] 
$. When $\mathbf{k}$ is real, the energy levels are pure spin states,
quantized along $\mathbf{\chi }$, in the plane perpendicular to $\mathbf{k}$%
. Note that the two eigenvalues of $\mathbf{\chi }\cdot \widehat{\mathbf{%
\sigma }}$\ are opposite, equal to the square roots of $\overline{\chi }%
^{2}=\chi _{x}^{2}+\chi _{y}^{2}+\chi _{z}^{2}$.\textbf{\ }We designate by $%
\overline{\chi }_{+}$\ $\left( \overline{\chi }_{-}\right) $ the square
roots of $\overline{\chi }^{2}$ ($\overline{\chi }_{+}$with a positive real
part and $\overline{\chi }_{-}$with a negative real part, if relevant). $%
\overline{\chi }_{+}\,\left( \overline{\chi }_{-}\right) $ will be used in
Eqs. \ref{psi_plu_mo_100} and \ref{psi_plu_mo_120}. The eigenvalues of $%
\widehat{H}$ are written as $\mathcal{E}\left( \mathbf{k}\right) $.

Inside a finite-width barrier, the incident plane wave $e^{iqz}$ is usually
to be replaced by $e^{\mp Kz}$ which corresponds to an imaginary wave vector 
$\pm iK$.

a) If the incident wave vector $\mathbf{k}_{I}$ is in the [001] direction $%
\left( \mathbf{k}_{I}=\left[ 0,\,0,\,q\right] \right) $, the wave vector in
the barrier is $\mathbf{k}_{II}=\left[ 0,\,0,\,\pm iK\right] $ and the
degenerate eigenvalues of $\widehat{H}$ are $\mathcal{E}\left( \mathbf{k}%
\right) =-\gamma _{c}K^{2}$ which is the (real) energy $E\left( \mathbf{k}%
\right) $ in the forbidden band gap. If $\mathbf{k}_{I}$ is almost in the
[001] direction $\left( \mathbf{k}_{I}=\left[ \xi ,\,0,\,q\right] \text{
with }\xi \ll q\right) $, $\mathbf{k}_{II}=\left[ \xi ,\,0,\,\pm iK\right] $
and the eigenvalues of $\widehat{H}$ are $\mathcal{E}\left( \mathbf{k}%
\right) =-\gamma _{c}\left( K^{2}-\xi ^{2}\right) \pm \gamma \xi K\sqrt{%
K^{2}-\xi ^{2}}$ which is the energy $E\left( \mathbf{k}\right) $ in the
forbidden band gap as well.

b) If $\mathbf{k}_{I}$ is in the [110] direction $\left( \mathbf{k}_{I}=%
\frac{q}{\sqrt{2}}\left[ 110\right] \right) $, a simple idea would be to
take $\mathbf{k}_{II}=\pm \frac{iK}{\sqrt{2}}\left[ 110\right] $ which leads
to $\mathcal{E}\left( \mathbf{k}\right) =-\gamma _{c}K^{2}\pm i\frac{\gamma 
}{2}K^{3}$. This quantity is not real and cannot be an energy $E\left( 
\mathbf{k}\right) $.\cite{imag}\ We are therefore led to consider a wave
vector such that $\mathbf{k}_{II}=\frac{1}{\sqrt{2}}\left( Q\pm iK\right) %
\left[ 110\right] $.

The calculation is given in Appendix A. The resulting band is plotted in
Fig. 4, over a very broad energy domain to reveal its general structure. We
are only interested in evanescent states located in the forbidden band gap,
i.e. states with a small negative energy. For our purposes, a key point is
that, at a given energy, we have \textit{exactly} the four possible states,
the wave vectors of which are $(Q\pm iK)$ with spin $\uparrow $ and $(-Q\pm
iK)$ with spin $\downarrow $, the latter being obtained from the former
through $\widehat{K}$. In short:%
\begin{equation}
E_{\uparrow }\left( \mathbf{k}\right) =E_{\uparrow }\left( \mathbf{k}^{\ast
}\right) =E_{\downarrow }\left( -\mathbf{k}\right) =E_{\downarrow }\left( -%
\mathbf{k}^{\ast }\right)  \label{Q_spin}
\end{equation}

Eq. \ref{Q_spin} provides us with a concrete example of the ideas developed
by Jones\cite{Jones} who showed that $E\left( \mathbf{k}\right) =E\left( 
\mathbf{k}^{\ast }\right) $. The corresponding four plane waves are $%
e^{i\left( Q\pm iK\right) }\uparrow $, $e^{i\left( -Q\pm iK\right)
}\downarrow $ or $e^{\mp Kz}e^{iQz}\uparrow $, $e^{\mp Kz}e^{-iQz}\downarrow 
$ (this is schematically shown in Fig. 5). This leads us to define 
\begin{equation}
\Uparrow =e^{iQz}\uparrow \qquad \Downarrow =e^{-iQz}\downarrow  \label{FHFB}
\end{equation}%
so that the four plane waves write $e^{\mp Kz}\Uparrow $ and $e^{\mp
Kz}\Downarrow $.

In the following, $\uparrow $ and $\downarrow $ are the up and down spins
when the $\mathbf{\chi }$ vector, which plays the role of a magnetic field,
lies along a real direction and is taken as quantization axis. When $\mathbf{%
\chi }$ is not collinear to any real direction, the spin eigenstates are $%
\uparrow _{\mathbf{k}}$ and $\downarrow _{\mathbf{k}}$: In Subsec. \ref{zero}%
, we shall see that $\uparrow _{\mathbf{k}}$ and $\downarrow _{\mathbf{k}}$
are no longer orthogonal. The implications of a wave vector $k=Q\pm iK$ in
the $[110]$ direction will be considered in detail in Sec. III.

\subsection{Probability current}

\subsubsection{The free-electron probability current}

We consider a spin-orbit-split barrier separating two regions where the
electron states are described by plane waves and where the potential is
taken equal to zero, as shown in Fig. 1. The barrier potential is assumed to
be a positive constant. When dealing with tunneling phenomena through
crystalline barriers, the wave vector component $\mathbf{\xi }$ parallel to
the barrier plane has to be conserved. For an incident plane wave, which has
a real wave vector component parallel to the surface plane, this implies
that the imaginary component of the wave vector inside the barrier has to be
orthogonal to the barrier plane. Then, the imaginary component of the wave
vector inside the barrier defines the tunneling direction. To analyze the
tunneling processes, we distinguish two different mechanisms: i) The wave
vector has collinear real and imaginary components along the normal to the
barrier (we refer to this mechanism as \textit{para}-type) and ii) The real
and imaginary components of the wave vector in the barrier are orthogonal
(we refer to this mechanism as \textit{ortho}-type). We would point out that
a plane wave with the real (imaginary) wave vector $\left( \mathbf{\xi +Q}%
\right) $ $($resp.$\ i\mathbf{K})$ is associated with the "classical"
probability current, e.g. calculated for a free electron, $\mathbf{J}%
^{f}=\hbar \left( \mathbf{\xi +Q}\right) /m$ $($resp.$\ \mathbf{0})$. Such
currents, with a zero divergence, conserve the local probability\textbf{\ }%
in any domain located in the barrier. On the contrary, a plane wave with the
wave vector $\left( \mathbf{\xi +Q}\right) +i\mathbf{K}$ is associated to $%
\mathbf{J}^{f}=e^{-2\mathbf{K\cdot r}}$~$\hbar \left( \mathbf{\xi +Q}\right)
/m$ : It looks as if the local probability were to be no longer conserved in
a domain located in the barrier, unless $\mathbf{Q=0}$, because $\mathbf{%
\nabla \cdot J}^{f}=-\left( 2\hbar /m\right) \left( \mathbf{K\cdot Q}\right)
e^{-2\mathbf{K\cdot r}}$. The case $Q=0$, results in a laminar free-electron
probability\textbf{\ }flux. The loops in the complex band structure which
have been studied in Ref. 
%TCIMACRO{\TeXButton{\onlinecite{Richard}}{\onlinecite{Richard}} }%
%BeginExpansion
\onlinecite{Richard}
%EndExpansion
correspond to ortho-tunneling, the normal to the barrier plane lying along
[001], a direction where the spin splitting is zero. On the contrary,
tunneling along the [110] direction,\textbf{\ }a direction where the DP
field is maximum\textbf{,} is a para-process. More precisely, the definition
of the free-electron current probability,%
\begin{equation}
\mathbf{J}^{f}\left[ \psi \right] =\func{Re}\left[ \psi ^{\ast }\dfrac{%
\widehat{\mathbf{p}}}{m}\psi \right] =\frac{\hbar }{m}\func{Im}\left[ \psi
^{\ast }\mathbf{\nabla }\psi \right]  \label{J_free}
\end{equation}%
is obtained from the conservation of the local probability when the
potential in the Schr\"{o}dinger equation is real.\cite{CTDL-proba}
Obviously, the equations expressing the conservation of the probability have
to be re-examined carefully in our case, where the Hermitian potential is
non-real due to the spin-orbit interaction. The detailed derivation of the
relevant current operator which allows one to calculate the true currents of
probability $\mathbf{J}_{\pm }$ and $\mathbf{J}=\mathbf{J}_{+}+\mathbf{J}%
_{-} $ is given in Appendix \ref{defi_curr}. There, it is shown how to
extend the usual procedure, which consists to define the velocity $\widehat{v%
}$ from the relation%
\begin{equation}
\widehat{v}=\frac{\partial \widehat{H}}{\partial p}  \label{velo_groupe}
\end{equation}

\subsubsection{The ortho- and para-processes\label{ortho-para}}

Coming back to the specific case of the GaAs-type barrier, let us derive a
few basic results and introduce some definitions.\textbf{\ }The orbital part
of the wave function of the conduction band is $S$ in usual Kane's notation%
\cite{Kane} and we write $\psi _{+}=S\uparrow \left( \mathbf{k}\right) $ and 
$\psi _{-}=S\downarrow \left( \mathbf{k}\right) $ where $\uparrow \left( 
\mathbf{k}\right) =\uparrow $ [see i) below] or $\uparrow \left( \mathbf{k}%
\right) =\uparrow _{\mathbf{k}}$ [see ii) below] and $\downarrow \left( 
\mathbf{k}\right) =\downarrow $ [i)] or $\downarrow \left( \mathbf{k}\right)
=\downarrow _{\mathbf{k}}$ [ii)]. The corresponding Schr\"{o}dinger equation
is: 
\begin{equation}
i\hbar \dfrac{\partial \psi _{\pm }}{\partial t}=\dfrac{\widehat{\mathbf{p}}%
^{2}}{2m}\psi _{\pm }+\gamma \overline{\chi }_{\pm }(\mathbf{k})\ \psi _{\pm
}  \label{psi_plu_mo_100}
\end{equation}

i) \textit{Ortho-process}. Let us assume that $\mathbf{k}_{II}=\mathbf{\xi }%
+i\mathbf{K}$ ( i.e. $\func{Re}\mathbf{k}_{II}.\func{Im}\mathbf{k}_{II}=0$,
with $\mathbf{\xi \cdot K}=0$)\textbf{\ }is a possible evanescent state.
Because $E=\dfrac{\hbar ^{2}}{2m}\left( \xi ^{2}-K^{2}\right) $ is real on a
real-energy line, the terms $\overline{\chi }_{\pm }(\mathbf{k})$
originating from the spin part of the Hamiltonian are also to be real. We
follow the usual procedure to derive the expression of the probability
current. $\uparrow _{\mathbf{k}}$ and $\downarrow _{\mathbf{k}}$ are no
longer orthogonal but in any case the real spin term disappears so that we
obtain%
\begin{equation}
\dfrac{\partial \left\vert \psi _{\pm }\right\vert ^{2}}{\partial t}=-%
\mathbf{\nabla }\cdot \mathbf{J}^{f}\left[ \psi _{\pm }\right]
\label{grad-J}
\end{equation}%
which is the usual relation for probability conservation. Care has to be
taken that the relation $\mathbf{\nabla }\cdot \mathbf{J}^{f}\left[ \psi
_{\pm }\right] =\mathbf{\nabla }\cdot \mathbf{J}\left[ \psi _{\pm }\right] $
does not mean that $\mathbf{J}^{f}\left[ \psi _{\pm }\right] =\mathbf{J}%
\left[ \psi _{\pm }\right] $. However, in such a case, a number of classical
results derived for free electrons will be recovered.

ii) \textit{Para-process}. In the case of one-dimensional tunneling along
the $\mathbf{n}$ direction, where $\mathbf{n}$ is a unit vector normal to
the barrier, which involves a complex wave vector $\mathbf{k}=\left(
Q+iK\right) \mathbf{n}$, $\mathbf{\chi }(\mathbf{k})=\left( Q+iK\right) ^{3}%
\mathbf{\chi }\left( \mathbf{n}\right) $, we quantize the spin along the
direction of $\mathbf{\chi }\left( \mathbf{n}\right) $ which is a real
non-normalized vector. $\overline{\chi }_{\pm }(\mathbf{k})$ are no longer
real. We follow the same procedure to derive the expression of the
probability current and we obtain%
\begin{equation}
\dfrac{\partial \left\vert \psi _{\pm }\right\vert ^{2}}{\partial t}=-%
\mathbf{\nabla }\cdot \mathbf{J}^{f}\left[ \psi _{\pm }\right] +\dfrac{2\ }{%
\mathbb{\hbar }}\gamma \func{Im}\overline{\chi }_{\pm }\ \left\vert \psi
_{\pm }\right\vert ^{2}  \label{psi_plu_mo_120}
\end{equation}

These equations could suggest an interpretation in terms of two-channel
transport with a generation-recombination rate, analogous to Giant
MagnetoResistance phenomena.\cite{Valet} In such a case, we would
classically expect a spin mixing and we will show that, indeed, a formal
analogy exists. However, care has to be taken that, at a given $\mathbf{k}$, 
$\psi _{+}$ and $\psi _{-}$ do not correspond to the same energy except when 
$\overline{\chi }$ is zero.

\subsubsection{The $\left[ 110\right] $ direction}

More specifically, we will deal with electron tunneling along the $[110]$
direction, a direction where the spin splitting is maximum in the real
conduction band. On this example, we illustrate the preceding
considerations. Let us consider for instance the up-spin channel\textbf{%
\textit{, }}where a possible wave vector is\textbf{\textit{\ }}$\mathbf{k}%
=\left( Q+iK\right) \mathbf{e}_{110}$ as shown in Sec. \ref{energie}, with
the wave function%
\begin{equation}
\psi _{+}(z)=e^{i\left( Q+iK\right) z}  \label{110_a}
\end{equation}%
and the DP field%
\begin{equation}
\overline{\chi }_{+}=\frac{1}{2}\left( Q+iK\right) ^{3}  \label{110_B}
\end{equation}%
The free-electron current is%
\begin{equation}
\mathbf{J}_{+}^{f}\left[ \psi _{+}\right] =\frac{\mathbb{\hbar }}{m}\func{Im}%
\psi _{+}^{\ast }\mathbf{\nabla }\psi _{+}=\frac{\mathbb{\hbar }Q}{m}e^{-2Kz}%
\text{ }  \label{110_d}
\end{equation}%
\begin{equation}
\mathbf{\nabla \cdot J}_{+}^{f}\left[ \psi _{+}\right] =-\frac{2\mathbb{%
\hbar }}{m}KQe^{-2Kz}  \label{110_e}
\end{equation}%
On a real-energy line (see Eq. \ref{Q_180}) 
\begin{eqnarray}
\dfrac{\partial \left\vert \psi _{+}\right\vert ^{2}}{\partial t} &=&-%
\mathbf{\nabla \cdot J}_{+}\left[ \psi _{+}\right] =\frac{2\mathbb{\hbar }}{m%
}KQe^{-2Kz}+\frac{\gamma }{\mathbb{\hbar }}\func{Im}\left( Q+iK\right)
^{3}e^{-2Kz}  \notag \\
&=&\frac{2K}{\mathbb{\hbar }}\left[ 2\gamma _{c}Q+\frac{1}{2}\gamma \left(
3Q^{2}-K^{2}\right) \right] e^{-2Kz}=0  \label{110_f}
\end{eqnarray}

Along the real-energy line, the eigenstates of the Schr\"{o}dinger equation
comply, as expected, with the continuity equation, the current $\mathbf{J}%
_{+}$ being to be identified. Here, it is easy to show that (see Appendix %
\ref{defi_curr}) 
\begin{equation}
\mathbf{J}_{\pm }\left[ \psi _{\pm }\right] =\mathbf{J}^{f}\left[ \psi _{\pm
}\right] \pm \frac{\gamma }{2\hbar }\left( 3\left\vert \frac{\partial }{%
\partial z}\psi _{\pm }\right\vert ^{2}-\frac{\partial ^{2}}{\partial z^{2}}%
\left\vert \psi _{\pm }\right\vert ^{2}\right)  \label{110_g}
\end{equation}

In the real conduction band, taking $\psi _{\pm }=e^{iqz}$, we obtain%
\begin{equation}
\mathbf{J}_{\pm }\left[ e^{iqz}\right] =\frac{2\mathbf{\gamma }_{c}}{\hbar }%
q\pm \frac{3}{2}\frac{\gamma }{\hbar }q^{2}=\frac{1}{\hbar }\frac{\partial }{%
\partial q}\left[ \gamma _{c}q^{2}\pm \frac{1}{2}\gamma q^{3}\right] =\frac{1%
}{\hbar }\frac{\partial }{\partial q}E\left( q\right)  \label{110_h}
\end{equation}

Concerning an evanescent wave $\psi _{\pm }=e^{\left( K\pm iQ\right) z}$, it
is easy to check that $\mathbf{J}_{\pm }\left[ e^{\left( K\pm iQ\right) z}%
\right] =0$ on a real-energy line.

\subsubsection{Waves conserving the free-electron probability current\label%
{J_subspace}}

The waves which conserve the free-electron current of probability play a
special role: They appear to be "quasi-classical states" which allow us to
build solutions yielding intuitive physical interpretations. The waves
involved in an ortho-process verify $\func{Re}\mathbf{k}\cdot \func{Im}%
\mathbf{k}=0$ and we have seen in Sec. \ref{ortho-para} that this condition
ensures the conservation of $\mathbf{J}^{f}$. In the case of a para-process,
a paradigm being tunneling along $[110]$, $\mathbf{J}^{f}$ is not conserved
in a given spin channel. Therefore, it is necessary to consider an
intricated wave function, $\psi \left( \mathbf{r}\right) =\psi _{+}\left( 
\mathbf{r}\right) \uparrow +\psi _{-}\left( \mathbf{r}\right) \downarrow
=\psi _{+}\uparrow +\psi _{-}\downarrow =\psi _{\uparrow }\uparrow +\psi
_{\downarrow }\downarrow $. In the following, we indifferently use the
notation $\psi _{+}$ and $\psi _{-}$ or $\psi _{\uparrow }$ and $\psi
_{\downarrow }$.

The free-electron probability current is given by\cite{Messiah} $\mathbf{J}%
^{f}\left[ \psi \right] =\left( 1/m\right) \func{Re}\left\langle \psi ^{\ast
}\widehat{\mathbf{p}}\psi \right\rangle _{\sigma }$ where the index $\sigma $
means a summation (partial trace) on the spin or $\mathbf{J}^{f}\left[ \psi %
\right] =\left( 1/m\right) \func{Re}\left( \psi _{+}^{\ast }\widehat{\mathbf{%
p}}\psi _{+}+\psi _{-}^{\ast }\widehat{\mathbf{p}}\psi _{-}\right) =\mathbf{J%
}^{f}\left[ \psi _{+}\right] +\mathbf{J}^{f}\left[ \psi _{-}\right] $. Due
to Kramers symmetry, the wave functions in the barrier $\psi _{II+}$ and $%
\psi _{II-}$ can be written%
\begin{eqnarray}
\psi _{II+}(z) &=&A_{2}e^{i\left( Q+iK\right) }+B_{2}e^{i\left( Q-iK\right)
z}  \notag \\
\psi _{II-}(z) &=&\tilde{A}_{2}e^{i\left( -Q+iK\right) }+\tilde{B}%
_{2}e^{i\left( -Q-iK\right) z}  \label{110_i}
\end{eqnarray}

The free-electron probability current carried by the function of the type $%
\phi =\left( A_{2}e^{-Kz}+B_{2}e^{Kz}\right) e^{i\epsilon Qz}$ is $\left(
\epsilon =\pm 1\right) $%
\begin{equation}
\mathbf{J}^{f}\left[ \phi \right] =\frac{\hbar }{m}\left[ 2K\func{Im}%
A_{2}^{\ast }B_{2}+\epsilon Q\left( 2\func{Re}A_{2}B_{2}^{\ast }+\left\vert
A_{2}\right\vert ^{2}e^{-2Kz}+\left\vert B_{2}\right\vert ^{2}e^{2Kz}\right) %
\right]  \label{J_phi}
\end{equation}

In the barrier, let us write $\Psi _{II}=\Psi _{II+}\uparrow +\Psi
_{II-}\downarrow $, so we have%
\begin{multline}
\mathbf{J}^{f}\left[ \Psi _{II}\right] \mathbf{=}\frac{\hbar }{m}\left\{ 2Q%
\left[ \func{Re}B_{2}A_{2}^{\ast }-\func{Re}\widetilde{B}_{2}\widetilde{A}%
_{2}^{\ast }\right] +2K\left[ \func{Im}B_{2}A_{2}^{\ast }+\func{Im}%
\widetilde{B}_{2}\widetilde{A}_{2}^{\ast }\right] \right. \\
\left. +Q\left[ e^{-2Kz}\left( \left\vert A_{2}\right\vert ^{2}-\left\vert 
\widetilde{A}_{2}\right\vert ^{2}\right) +e^{2Kz}\left( \left\vert
B_{2}\right\vert ^{2}-\left\vert \widetilde{B}_{2}\right\vert ^{2}\right) %
\right] \right\}  \label{J_PSI_II}
\end{multline}

We see that the probability current in the barrier is constant if and only
if $\left\vert A_{2}\right\vert =\left\vert \widetilde{A}_{2}\right\vert $, $%
\left\vert B_{2}\right\vert =\left\vert \widetilde{B}_{2}\right\vert $. This
leads to $A_{2}=\mathcal{A}e^{i\theta _{A}}$, $\widetilde{A}_{2}=\mathcal{A}%
e^{-i\theta _{A}}$, $B_{2}=\mathcal{B}e^{i\theta _{B}}$, and $\widetilde{B}%
_{2}=\mathcal{B}e^{-i\theta _{B}}$ where $\mathcal{A}$ and $\mathcal{B}$ are
two complex numbers. So the general expression of a wave \textit{sustaining
a constant }$\mathbf{J}^{f}$ inside the barrier is%
\begin{equation}
\Psi _{II}=\Psi _{II}(z)=\mathcal{A}e^{-Kz}\left[ e^{i\theta _{A}}\Uparrow
+e^{-i\theta _{A}}\Downarrow \right] +\mathcal{B}e^{Kz}\left[ e^{i\theta
_{B}}\Uparrow +e^{-i\theta _{B}}\Downarrow \right]  \label{PsiII_AB}
\end{equation}

It is useful to write 
\begin{equation}
\Psi _{II}(z)=\mathcal{A}e^{-Kz}\,S_{\exp i\theta _{A}}+\mathcal{B}%
e^{Kz}\,S_{\exp i\theta _{B}}  \label{Psi-II-bis}
\end{equation}%
where 
\begin{equation}
S_{\lambda }=S_{\lambda }\left( z\right) =\lambda \Uparrow +\lambda ^{\ast
}\Downarrow  \label{S_FHFB_ter}
\end{equation}

The Kramers conjugate of $S_{\lambda }$ is $\widehat{S}_{\lambda }=\widehat{K%
}S_{\lambda }$ where $\widehat{K}$ is the time reversal transformation.
Observe that $S_{\lambda }$ and $\widehat{S}_{\lambda }$ are eigenstates of
the helicity operator $\left( \widehat{\mathbf{p}}\cdot \widehat{\mathbf{%
\sigma }}\right) $ for the eigenvalue $\hbar Q$.

Let us look at the spin direction defined by $S_{\lambda }$. Recall that the
spin quantization direction is along the $\mathbf{\chi }\left( \mathbf{e}%
_{110}\right) $ vector. We call $Oz^{\prime }$ the direction parallel to $%
\mathbf{\chi }\left( \mathbf{e}_{110}\right) $; $Ox^{\prime }$ and $%
Oy^{\prime }$ are in the $\mathbf{\Pi }_{\mathbf{\chi }}$ plane normal to $%
\mathbf{\chi }\left( \mathbf{e}_{110}\right) $.\ The spin direction is
defined via $\left\langle \sigma _{x^{\prime }}\right\rangle $, $%
\left\langle \sigma _{y^{\prime }}\right\rangle $, $\left\langle \sigma
_{z^{\prime }}\right\rangle $. First of all we note that $\left\langle
\sigma _{z^{\prime }}\right\rangle =0$ while $\left\langle \sigma
_{x^{\prime }}\right\rangle =2\func{Re}\lambda ^{2}$ and $\left\langle
\sigma _{y^{\prime }}\right\rangle =-2\func{Im}\lambda ^{2}$ for $S_{\lambda
}\left( 0\right) $. The spin is in the $\mathbf{\Pi }_{\mathbf{\chi }}$
plane. Any spin direction in the $\mathbf{\Pi }_{\mathbf{\chi }}$ plane,
that\ we call an \textit{in-plane} direction, can be described by a suited
value of $\lambda $. For instance with $\lambda =\exp i\theta _{\lambda }$, $%
\left\langle \sigma _{x^{\prime }}\right\rangle =\cos 2\theta _{\lambda }$,
and $\left\langle \sigma _{y^{\prime }}\right\rangle =-\sin 2\theta
_{\lambda }$,\textbf{\ }apart a common factor\textbf{, }$\theta _{\lambda }$
being the angle between the\textbf{\textit{\ }}$Ox^{\prime }$ axis and the
spin direction.

It can be shown that the largest vectorial space consisting of $\mathbf{J}%
^{f}$-conserving waves at a given energy is $\mathfrak{E=}\left\{ \Psi
_{\alpha ,\,\beta }\right\} $ where%
\begin{equation}
\Psi _{\alpha ,\,\beta }=\left( \alpha \mathcal{A}e^{-Kz}+\beta \mathcal{B}%
e^{Kz}\right) \Uparrow +\left( \alpha ^{\ast }\mathcal{A}e^{-Kz}+\beta
^{\ast }\mathcal{B}e^{Kz}\right) \Downarrow  \label{HJD_100}
\end{equation}%
with $\alpha $ and $\beta \in \mathbb{C}$. $\mathfrak{E}$ is a vectorial
space over $%
%TCIMACRO{\U{211d} }%
%BeginExpansion
\mathbb{R}
%EndExpansion
$, but not over $%
%TCIMACRO{\U{2102} }%
%BeginExpansion
\mathbb{C}
%EndExpansion
$.

Moreover, the existence of a superposition principle implies that any linear
combination with real coefficients of two solutions with a current of
probability of a given sign has to be a solution associated to a current of
probability of the same sign. This is a strong constraint which is verified
over $\mathfrak{E}_{0}\mathfrak{=}\left\{ \Phi _{\mathcal{A},\,\mathcal{B}%
}\right\} \otimes \left\{ S_{\alpha }\right\} =\left\{ \left( \mathcal{A}%
e^{-Kz}+\mathcal{B}e^{Kz}\right) S_{\alpha }\right\} $, a vectorial subspace
of $\mathfrak{E}$ (in this subspace $\mathbf{J}^{f}\left[ \Phi _{\mathcal{A}%
,\,\mathcal{B}}S_{\alpha }\right] =2\left\vert \alpha \right\vert ^{2}%
\mathbf{J}^{f}\left[ \Phi _{\mathcal{A},\,\mathcal{B}}\right] $), or in $%
\left\{ (\Phi _{\mathcal{A},\,\mathcal{B}}S\mathbb{)}_{\lambda ,\,\theta
}=\cos \theta \,\Phi _{\mathcal{A},\,\mathcal{B}}\,S_{\lambda }+\sin \theta
\,\frac{1}{K}\,\frac{\partial }{\partial z}\Phi _{\mathcal{A},\,\mathcal{B}%
}\left( i\widehat{S}_{\lambda }\right) \right\} _{\theta }$- at fixed $%
\theta $ - which also is vectorial subspace (in this subspace $\mathbf{J}^{f}%
\left[ (\Phi _{\mathcal{A},\,\mathcal{B}}S\mathbb{)}_{\lambda ,\,\theta }%
\right] =2\left\vert \alpha \right\vert ^{2}\mathbf{J}^{f}\left[ \Phi _{%
\mathcal{A},\,\mathcal{B}}\right] \cos 2\theta $).

\subsection{Standard tunneling case\label{standard}}

The standard tunneling case is to be recovered when $\gamma $ is zero,
therefore, we build our analysis in close relation with it. A crucial point
is that the probability current has to be constant so that $R+T=1$ where $R$ 
$\left( T\right) $ is the reflection (transmission) coefficient.

We shall need the standard (without spin) function $\psi ^{\left( 0\right)
}\left( z\right) $ defined as:%
\begin{equation}
\psi ^{\left( 0\right) }\left( z\right) =\left\{ 
\begin{array}{llll}
\psi _{I}^{\left( 0\right) }\left( z\right) & = & a_{1}e^{iqz}+b_{1}e^{-iqz}
& \quad \left( z<0\right) \\ 
\psi _{II}^{\left( 0\right) }\left( z\right) & = & a_{2}e^{-Kz}+b_{2}e^{Kz}
& \quad \left( 0<z<a\right) \\ 
\psi _{III}^{\left( 0\right) }\left( z\right) & = & a_{3}e^{iqz} & \quad
\left( a<z\right)%
\end{array}%
\right.  \label{PHI-100}
\end{equation}%
where $z<0$, $0<z<a$, and $a<z$ respectively correspond to the incident wave
(index $I$), to the wave in the barrier (index $II$), and to the transmitted
wave (index $III$), as illustrated in Fig. 1.

$\psi ^{\left( 0\right) }\left( z\right) $, a $C^{1}$-function, meets the
boundary conditions 
\begin{equation}
\psi ^{\left( 0\right) }\left( z_{0-}\right) =\psi ^{\left( 0\right) }\left(
z_{0+}\right) \text{, }\frac{\partial \psi ^{0}\left( z_{0-}\right) }{%
\partial z}=\frac{\partial \psi ^{\left( 0\right) }\left( z_{0+}\right) }{%
\partial z}\text{, }z_{0}=0\text{ or }a  \label{PHI-120}
\end{equation}

\begin{align}
\frac{b_{1}}{a_{1}}& =\frac{2\left( q^{2}+K^{2}\right) \sinh Ka}{D}\overset{%
\exp Ka\gg 1}{\approx }\frac{\left( q^{2}+K^{2}\right) }{\left( q+iK\right)
^{2}}  \notag \\
\frac{a_{2}}{a_{1}}& =\frac{2q\left( q+iK\right) e^{Ka}}{D}\overset{\exp
Ka\gg 1}{\approx }2\frac{q}{(q+iK)}  \notag \\
\frac{b_{2}}{a_{1}}& =\frac{2q\left( -q+iK\right) e^{-Ka}}{D}\overset{\exp
Ka\gg 1}{\approx }2\frac{q\left( -q+iK\right) }{(q+iK)^{2}}e^{-2Ka}
\label{ab_total} \\
\frac{a_{3}}{a_{1}}& =\frac{4iKq}{D}e^{-iqa}\overset{\exp Ka\gg 1}{\approx }%
4i\dfrac{qKe^{-iqa}}{(q+iK)^{2}}e^{-Ka}  \notag \\
D& =\left( q+iK\right) ^{2}e^{Ka}-\left( q-iK\right) ^{2}e^{-Ka}  \notag
\end{align}

The function $\psi ^{\left( 0\right) }\left( z\right) $ is such that the
probability current $\mathbf{J}^{f}\left[ \psi ^{\left( 0\right) }\right] $
is constant. The reflection coefficient $R=\left\vert b_{1}/a_{1}\right\vert
^{2}$ and the transmission coefficient $T=\left\vert a_{3}/a_{1}\right\vert
^{2}$ are such that $R+T=1$.

Also observe that, if we multiply $\psi ^{\left( 0\right) }$ by any $C^{1}$%
-function $f(\mathbf{r},\,\uparrow ,\,\downarrow )$, the new function and
its derivative are continuous at the interfaces, satisfying the initial
boundary conditions. Consider the case where\ the incident wave is $e^{i%
\mathbf{q\cdot r}}$.\ If we take $f(\mathbf{r},\,\uparrow ,\,\downarrow
)=e^{i\mathbf{\xi \cdot r}}\uparrow $ or $f(\mathbf{r},\,\uparrow
,\,\downarrow )=e^{i\mathbf{\xi \cdot r}}\downarrow $, we obtain a solution
to the tunneling problem if, and only if, the incident component $e^{i\left( 
\mathbf{q+\xi }\right) \cdot \mathbf{r}}$ and the reflected component $%
e^{i\left( -\mathbf{q+\xi }\right) \cdot \mathbf{r}}$, correspond to the
same energy.\cite{noteSi}

\section{Para process: [110]-oriented barrier under normal incidence}

\subsection{General considerations}

In the case where the wave vector is parallel to the [110] direction and the 
$\widehat{H}_{DP}$ Hamiltonian is taken into account, we have seen in Sec. %
\ref{energie} that the wave vector is to be of the form $\left( \epsilon
Q\pm iK\right) \mathbf{e}_{110}$ to get a real eigenvalue (an energy) of the
Hamiltonian. But in such case, $\mathbf{J}^{f}$ is not conserved (Eq. \ref%
{J_phi}) and even not constant inside the barrier, so that the standard
calculation routine to find the solution (i.e. the continuity of the wave
function and of its derivative) cannot apply.

We could try to build a solution according to the usual procedure, but with
a wave in the barrier involving\ the two spin channels, which can give a
constant $\mathbf{J}^{f}$ (see Eq. \ref{PsiII_AB}) 
\begin{equation}
\Psi \left( z\right) =\left\{ 
\begin{array}{lll}
\Psi _{I}\left( z\right) & = & \left( A_{1}e^{iqz}+B_{1}e^{-iqz}\right)
\uparrow +\widetilde{B}_{1}e^{-iqz}\downarrow \\ 
\Psi _{II}\left( z\right) & = & \left( A_{2}e^{-Kz}+B_{2}e^{Kz}\right)
\Uparrow +\left( \widetilde{A}_{2}e^{-Kz}+\widetilde{B}_{2}e^{Kz}\right)
\Downarrow \\ 
\Psi _{III}\left( z\right) & = & A_{3}e^{iqz}\uparrow +\widetilde{A}%
_{3}e^{iqz}\downarrow%
\end{array}%
\right.  \label{PSI-100}
\end{equation}%
where $\Uparrow $ and $\Downarrow $ are defined in Eq. \ref{FHFB}.

The usual boundary conditions ($C^{1}$ function) for the down-spin channel
for instance yield four equations determining $\widetilde{B}_{1}$, $%
\widetilde{A}_{2}$, $\widetilde{B}_{2}$, and $\widetilde{A}_{3}$ 
\begin{eqnarray}
\widetilde{B}_{1} &=&\widetilde{B}_{2}+\widetilde{A}_{2}  \notag \\
q\widetilde{B}_{1} &=&\left( Q-iK\right) \widetilde{A}_{2}+\left(
Q+iK\right) \widetilde{B}_{2}  \notag \\
\widetilde{A}_{2}e^{-i\left( Q-iK\right) a}+\widetilde{B}_{2}e^{-i\left(
Q+iK\right) a} &=&\widetilde{A}_{3}e^{iqa}  \notag \\
\widetilde{A}_{2}\left( Q-iK\right) e^{-i\left( Q-iK\right) a}+\widetilde{B}%
_{2}\left( Q+iK\right) e^{-i\left( Q+iK\right) a} &=&-\widetilde{A}%
_{3}qe^{iqa}  \label{PSI-110}
\end{eqnarray}

They only provide a non trivial solution if the determinant of the system is
equal to $0$ which gives the relation%
\begin{equation}
\left( q^{2}-Q^{2}-K^{2}\right) \sinh Ka+2iKq\cosh Ka=0  \label{E_qQK}
\end{equation}

The only solution is $K=0$ but it is not relevant to our problem.

\subsection{Solutions to the tunneling problem\label{Solutions}}

\subsubsection{The constant-$\protect\gamma $ case}

We go back to the Schr\"{o}dinger equation to determine the proper boundary
conditions and, to avoid any unnecessary mathematical complexity, we here
assume that $\gamma $ is constant over the three regions. Along the $[110]$
direction, with $\mathbf{k}=\left( 1/\sqrt{2}\right) k[110]$, the DP
Hamiltonian writes%
\begin{equation}
H_{DP}=\gamma _{c}k^{2}\pm \frac{1}{2}\gamma k^{3}  \label{H_ud1}
\end{equation}%
where the $+$ (resp. $-$) sign applies to the up (down) spin, quantized
along the DP field. As usual, we obtain the effective Hamiltonian by
substituting $\mathbf{k}$ with $-i\mathbf{\nabla }$ , i.e. $k$ with $-i%
\dfrac{\partial }{\partial z}$.%
\begin{equation}
H_{DP}=-\gamma _{c}\frac{\partial ^{2}}{\partial z^{2}}\pm \frac{i}{2}\gamma 
\frac{\partial ^{3}}{\partial z^{3}}\text{\textbf{\ }}  \label{H_ud2}
\end{equation}

Thus, we have the two equations%
\begin{eqnarray}
\left[ -\gamma _{c}\frac{\partial ^{2}}{\partial z^{2}}+\frac{1}{2}i\gamma 
\frac{\partial ^{3}}{\partial z^{3}}\right] \psi _{\uparrow } &=&\left[
E-V\left( z\right) \right] \psi _{\uparrow }\text{ }  \notag \\
\left[ -\gamma _{c}\frac{\partial ^{2}}{\partial z^{2}}-\frac{1}{2}i\gamma 
\frac{\partial ^{3}}{\partial z^{3}}\right] \psi _{\downarrow } &=&\left[
E-V\left( z\right) \right] \psi _{\downarrow }\text{ }  \label{H_ud_3}
\end{eqnarray}%
where $V\left( z\right) =V$ when $0\leq z\leq a$ and $V\left( z\right) =0$
outside. Because the DP Hamiltonian was obtained using the perturbation
theory, we will look for a solution to the effective Schr\"{o}dinger
equation to the first order in $\gamma $ only. Let us consider the up-spin
channel: We write%
\begin{equation}
\psi _{\uparrow }=\psi ^{\left( 0\right) }+\psi _{\uparrow }^{\left(
1\right) }  \label{solu_1}
\end{equation}%
where\ $\psi ^{\left( 0\right) }$ is the standard function (obtained for $%
\gamma =0$, and\textbf{\textit{\ }}defined by Eq. \ref{PHI-100}; It is a $%
C^{1}$ function, with a discontinuous second derivative). $\psi _{\uparrow
}^{\left( 1\right) }$ is a first-order term in $\gamma $ so that the Schr%
\"{o}dinger equation to the first order writes%
\begin{equation}
-\gamma _{c}\frac{\partial ^{2}\psi _{\uparrow }}{\partial z^{2}}+\frac{1}{2}%
i\gamma \frac{\partial ^{3}\psi ^{\left( 0\right) }}{\partial z^{3}}=\left[
E-V\left( z\right) \right] \psi _{\uparrow }  \label{Schro_up_first}
\end{equation}

We integrate this equation from one side of the interface to the other%
\begin{equation}
-\gamma _{c}\left[ \frac{\partial \psi _{\uparrow }}{\partial z}\right]
_{z_{0}-\varepsilon }^{z_{0}+\varepsilon }+\frac{1}{2}i\left[ \gamma \left( 
\frac{\partial ^{2}\psi _{\uparrow }^{(0)}}{\partial z^{2}}\right) \right]
_{z_{0}-\varepsilon }^{z_{0}+\varepsilon }=\int\limits_{z_{0}-\varepsilon
}^{z_{0}+\varepsilon }\left[ E-V\left( z\right) \right] \psi _{\uparrow }dz
\label{Schro_inter}
\end{equation}

Then%
\begin{equation}
\lim_{\varepsilon \longrightarrow 0}\left\{ -\gamma _{c}\left[ \frac{%
\partial \psi _{\uparrow }}{\partial z}\right] _{z_{0}-\varepsilon
}^{z_{0}+\varepsilon }+\frac{1}{2}i\left[ \gamma \left( \frac{\partial
^{2}\psi ^{\left( 0\right) }}{\partial z^{2}}\right) \right]
_{z_{0}-\varepsilon }^{z_{0}+\varepsilon }\right\} =0  \label{Bord4}
\end{equation}

Taking the standard function (Eq. \ref{PHI-100}\textbf{\ }) and referring to
the limit at $z_{0}$ inside the barrier and inside the well respectively as $%
z_{0}^{B}$ and $z_{0}^{W}$, we obtain 
\begin{equation}
\left[ \frac{\partial ^{2}\psi ^{\left( 0\right) }}{\partial z^{2}}\right]
_{z_{0}^{W}}=-q^{2}\psi ^{\left( 0\right) }(z_{0}^{W})  \label{deri_W}
\end{equation}%
outside the barrier, and%
\begin{equation}
\left[ \frac{\partial ^{2}\psi ^{\left( 0\right) }}{\partial z^{2}}\right]
_{z_{0}^{B}}=K^{2}\psi ^{\left( 0\right) }(z_{0}^{B})  \label{deri_B}
\end{equation}%
inside the barrier. At the interfaces $\psi ^{\left( 0\right)
}(z_{0}^{B})=\psi ^{\left( 0\right) }(z_{0}^{W})=\psi ^{\left( 0\right)
}(z_{0})$, then%
\begin{equation}
\left[ \frac{1}{2}i\gamma \left( \frac{\partial ^{2}\psi _{\uparrow }}{%
\partial z^{2}}\right) \right] _{z_{0}^{W}}^{z_{0}^{B}}=\frac{1}{2}i\gamma
\left( K^{2}+q^{2}\right) \psi ^{\left( 0\right) }(z_{0})  \label{Bord5}
\end{equation}

This provides us with the jump of the derivative at the interfaces. To the
first order in $q/K$%
\begin{equation}
\left[ \frac{\partial \psi _{\uparrow }}{\partial z}\right]
_{z_{0}^{W}}^{z_{0}^{B}}=\frac{1}{2}i\frac{\gamma }{\gamma _{c}}\left(
K^{2}+q^{2}\right) \psi ^{\left( 0\right) }\left( z_{0}\right) \thickapprox 
\frac{1}{2}i\frac{\gamma }{\gamma _{c}}K^{2}\psi ^{\left( 0\right) }\left(
z_{0}\right) =2iQ_{\uparrow }\psi ^{\left( 0\right) }\left( z_{0}\right)
\label{Bord6}
\end{equation}%
after Eq. \ref{Q_250}.

Similarly, for a down spin $Q_{\downarrow }=-Q_{\uparrow }$, and we have%
\begin{equation}
\left[ \frac{\partial \psi _{\downarrow }}{\partial z}\right]
_{z_{0}^{W}}^{z_{0}^{B}}=2iQ_{\downarrow }\psi ^{\left( 0\right) }\left(
z_{0}\right)  \label{Bord_down}
\end{equation}

It is worth remarking that this very\ discontinuity condition was found in a
quite different situation, involving Rashba-split quantum wells.\cite%
{Andrada}

Now let us assume that $Q_{\uparrow }=Q$. The wave function constructed from
the eigenstates in the three regions is%
\begin{equation}
\psi \left( z\right) =\left\{ 
\begin{array}{llll}
\psi _{I}\left( z\right) & = & A_{1}e^{iqz}+B_{1}(q,K,Q)e^{-iqz} & \quad
\left( z<0\right) \\ 
\psi _{II}\left( z\right) & = & 
A_{2}(q,K,Q)e^{-Kz}e^{iQz}+B_{2}(q,K,Q)e^{Kz}e^{iQz} & \quad \left(
0<z<a\right) \\ 
\psi _{III}\left( z\right) & = & A_{3}(q,K,Q)e^{iqz} & \quad \left(
a<z\right)%
\end{array}%
\right.  \label{Solu}
\end{equation}%
with the coefficients $B_{1}(q,K,Q),$ $A_{2}(q,K,Q),$ $B_{2}(q,K,Q),$ and $%
A_{3}(q,K,Q)$ to be determined.

To the first order in $Q$, the solution can be expanded as%
\begin{eqnarray}
\psi _{I}\left( z\right) &=&\left( a_{1}e^{iqz}+b_{1}e^{-iqz}\right) +\beta
_{1}Qe^{-iqz}  \notag \\
\psi _{II}(z) &=&\left( a_{2}e^{-Kz}+b_{2}e^{Kz}\right) e^{iQz}+Q\left(
\alpha _{2}e^{-Kz}+\beta _{2}e^{Kz}\right) e^{iQz}  \label{Solu_2} \\
\psi _{III}(z) &=&a_{3}e^{iqz}e^{iQa}+\alpha _{3}^{\prime }Qe^{iqz}  \notag
\end{eqnarray}%
with%
\begin{equation}
\beta _{1}=\left[ \frac{dB_{1}(q,K,Q)}{dQ}\right] _{Q=0}  \label{coeff_beta1}
\end{equation}%
\begin{equation}
\alpha _{2}=\left[ \frac{dA_{2}(q,K,Q)}{dQ}\right] _{Q=0}\ \ \text{;}\ \
\beta _{2}=\left[ \frac{dB_{2}(q,K,Q)}{dQ}\right] _{Q=0}\ 
\label{coeff_al2be2}
\end{equation}%
and%
\begin{equation}
\alpha _{3}=\left[ \frac{dA_{3}(q,K,Q)}{dQ}\right] _{Q=0}\simeq
iaa_{3}+\alpha _{3}^{\prime }  \label{coeff_al3p}
\end{equation}

We write%
\begin{equation}
\psi =\varphi ^{S}+\varphi ^{\widehat{S}}  \label{Solu_3}
\end{equation}%
where%
\begin{equation}
\varphi ^{S}\left( z\right) =\left\{ 
\begin{array}{llll}
\varphi _{I}^{S}\left( z\right) & = & a_{1}e^{iqz}+b_{1}e^{-iqz} & \quad
\left( z<0\right) \\ 
\varphi _{II}^{S}\left( z\right) & = & \left(
a_{2}e^{-Kz}+b_{2}e^{Kz}\right) e^{iQz} & \quad \left( 0<z<a\right) \\ 
\varphi _{III}^{S}\left( z\right) & = & a_{3}e^{iqz}e^{iQa} & \quad \left(
a<z\right)%
\end{array}%
\right.  \label{phiS}
\end{equation}%
\begin{equation}
\varphi ^{\widehat{S}}\left( z\right) =\left\{ 
\begin{array}{llll}
\varphi _{I}^{\widehat{S}}\left( z\right) & = & \beta _{1}Qe^{-iqz} & \quad
\left( z<0\right) \\ 
\varphi _{II}^{\widehat{S}}\left( z\right) & = & Q\left( \alpha
_{2}e^{-Kz}+\beta _{2}e^{Kz}\right) e^{iQz} & \quad \left( 0<z<a\right) \\ 
\varphi _{III}^{\widehat{S}}\left( z\right) & = & \alpha _{3}^{\prime
}Qe^{iqz} & \quad \left( a<z\right)%
\end{array}%
\right.  \label{phiSchap}
\end{equation}

$\varphi ^{S}$ is a continuous function but its derivative is not. To the
first order, its jump at the interfaces is%
\begin{equation}
\left[ \frac{\partial \varphi ^{S}}{\partial z}\right]
_{z_{0}^{W}}^{z_{0}^{B}}=iQ\psi _{II}^{\left( 0\right) }(z_{0})
\label{saut_phiS}
\end{equation}

As we have derived that the jump of the derivative of the wave function $%
\psi $ is $2iQ\psi _{II}^{\left( 0\right) }(z_{0})$, we deduce that $\varphi
^{\widehat{S}}$ is a continuous function and that the jump of its derivative
at the interfaces is%
\begin{equation}
\left[ \frac{\partial \varphi ^{\widehat{S}}}{\partial z}\right]
_{z_{0}^{W}}^{z_{0}^{B}}=iQ\psi _{II}^{\left( 0\right) }(z_{0})
\label{saut_phiSchap}
\end{equation}

This provides us with the four equations which determine the four
coefficients $\beta _{1},$ $\alpha _{2},$ $\beta _{2},$ and $\alpha
_{3}^{\prime }$%
\begin{eqnarray}
\beta _{1}-\alpha _{2}-\beta _{2} &=&0  \notag \\
\alpha _{2}e^{-Ka}+\beta _{2}e^{Ka}-\alpha _{3}^{\prime }e^{iqa} &=&0  \notag
\\
iq\beta _{1}-K\alpha _{2}+K\beta _{2} &=&i\psi _{II}^{\left( 0\right) }(0) 
\notag \\
K\alpha _{2}e^{-Ka}-K\beta _{2}e^{Ka}+iq\alpha _{3}^{\prime }e^{iqa}
&=&-i\psi _{II}^{\left( 0\right) }(a)  \label{System2}
\end{eqnarray}

The solution of this system is%
\begin{eqnarray}
\beta _{1} &=&-\frac{i}{K}a_{3}e^{iqa}\sinh aK=\frac{4q}{D}a_{1}\sinh aK 
\notag \\
\alpha _{2} &=&-ia_{3}e^{iqa}\frac{e^{Ka}}{2K}=\frac{a_{2}}{q+iK}  \notag \\
\beta _{2} &=&ia_{3}e^{iqa}\frac{e^{-Ka}}{2K}=\frac{b_{2}}{q-iK}  \notag \\
\alpha _{3}^{\prime } &=&0  \label{Solu_al_bet}
\end{eqnarray}

In the following, we consider $Q/K=\gamma K/4\gamma _{c}$ but also $q/K$ as
first order terms and we look for solutions up to the first order. The term
in the reflected wave function arising from $Q\beta _{1}$ is a second-order
contribution which has to be neglected. Note that, in Region $I$, if the
incident wave has the wave vector $q$, the reflected wave should have the
wave vector$\ -q^{\prime }$, where $q=q_{0}-\delta q$ and $q^{\prime
}=q_{0}+\delta q$. From Eq. \ref{Ener_well}, it can be verified that $\delta
q=\left( \gamma /4\gamma _{c}\right) q_{0}^{2}$ $=\left( q_{0}/K\right)
^{2}Q $. Then $\delta q$ is a second-order term which has to be neglected so
that media $I$ and $III$ have no sizable spin splitting. This indicates that
the solution we obtain in the case of a constant $\gamma $ also constitutes
a plausible physical solution when $\gamma $ is a step function, with $%
\gamma =0$ outside the barrier. Also note that, at this level of
approximation, $\varphi ^{\widehat{S}}$ is a wave which only exists inside
the barrier and is not coupled to the free-electron waves outside the
barrier.\ Because $A_{3}=a_{3}e^{iQa}$, we see that there is a pure
dephasing between the up- ($Q_{\uparrow }=Q$) and the down- ($Q_{\downarrow
}=-Q$) spin channels.

We have to be sure that, in our treatment, the probability current is
conserved along the tunnel process. The wave in the barrier, in the up-spin
channel, is of the form $\psi \left( z\right) =\left(
A_{2}e^{-Kz}+B_{2}e^{Kz}\right) e^{iQz}=\phi \left( z\right) e^{iQz}$ with $%
A_{2}=a_{2}\left( 1-i\frac{Q}{K}\right) $ and $B_{2}=b_{2}\left( 1+i\frac{Q}{%
K}\right) $. Let us calculate $\mathbf{J}$ to the first order in $Q$ by
making use of Eq. \ref{110_g}%
\begin{equation}
\mathbf{J}\left[ \psi \right] \mathbb{=}\mathbf{J}_{+}\left[ \psi \right] =%
\mathbf{J}^{f}\left[ \psi \right] +\frac{\gamma }{2\hbar }\left[ 3\left\vert 
\frac{\partial }{\partial z}\psi \right\vert ^{2}-\frac{\partial ^{2}}{%
\partial z^{2}}\left\vert \psi \right\vert ^{2}\right]  \label{110_g_bis}
\end{equation}

It is sufficient to evaluate the term in the bracket to the zeroth order,
substituting $\psi $ with $\psi ^{(0)}$. One finds%
\begin{eqnarray}
\mathbf{J}\left[ \psi \right] &\mathbb{=}&\mathbf{J}^{f}\left[ \psi \right] +%
\frac{\gamma }{2\hbar }\left[ \left\vert \frac{\partial \psi ^{\left(
0\right) }}{\partial z}\right\vert ^{2}-\psi ^{\left( 0\right) \ast }\frac{%
\partial ^{2}}{\partial z^{2}}\psi ^{\left( 0\right) }-\psi ^{\left(
0\right) }\frac{\partial ^{2}}{\partial z^{2}}\psi ^{\left( 0\right) \ast }%
\right]  \notag \\
&\thickapprox &\mathbf{J}^{f}\left[ \psi \right] +2\frac{\gamma _{c}}{\hbar K%
}\frac{Q}{K}\left[ \left\vert \frac{\partial \psi ^{\left( 0\right) }}{%
\partial z}\right\vert ^{2}-2K^{2}\left\vert \psi ^{\left( 0\right)
}\right\vert ^{2}\right]  \label{110_j} \\
&=&\mathbf{J}^{f}\left[ \psi \right] -2\frac{\gamma _{c}}{\hbar }Q\left[
\left\vert \psi ^{\left( 0\right) }\right\vert ^{2}+2\left( a_{2}^{\ast
}b_{2}+a_{2}b_{2}^{\ast }\right) \right]  \notag
\end{eqnarray}%
with%
\begin{equation}
\mathbf{J}^{f}\left[ \psi \right] =\func{Im}\left( \psi ^{\ast }\frac{\hbar 
}{m}\frac{\partial \psi }{\partial z}\right) =\frac{2\gamma _{c}}{\hbar }%
\func{Im}\left( \phi ^{\ast }\frac{\partial \phi }{\partial z}\right) +\frac{%
2\gamma _{c}}{\hbar }Q\left\vert \psi ^{\left( 0\right) }\right\vert ^{2}
\label{110_k}
\end{equation}%
\begin{equation}
\mathbf{J}^{f}\left[ \psi \right] =\frac{2\gamma _{c}}{\hbar }\func{Im}%
\left( \psi ^{\left( 0\right) \ast }\frac{\partial \psi ^{\left( 0\right) }}{%
\partial z}\right) +\frac{4\gamma _{c}}{\hbar }Q\left( a_{2}^{\ast
}b_{2}+a_{2}b_{2}^{\ast }\right) +\frac{2\gamma _{c}}{\hbar }Q\left\vert
\psi ^{\left( 0\right) }\right\vert ^{2}  \label{110_l}
\end{equation}

By comparing these expressions, one obtains%
\begin{equation}
\mathbf{J}\left[ \psi \right] =\frac{2\gamma _{c}}{\hbar }\func{Im}\left(
\psi ^{\left( 0\right) \ast }\frac{\partial \psi ^{\left( 0\right) }}{%
\partial z}\right) =\mathbf{J}^{f}\left[ \psi ^{\left( 0\right) }\right]
\label{110_m}
\end{equation}

This definitely establishes current conservation in the tunnel process.

Starting with an incident state $\left\vert \varphi _{0}\right\rangle $, the
transmission asymmetry $\mathcal{T}$ in the spin-dependent tunneling process
as can be expressed as%
\begin{equation}
\mathcal{T}=\frac{\left\vert \left\vert T\left( \left\vert \varphi
_{0}\right\rangle \right) \right\vert \right\vert ^{2}-\left\vert \left\vert
T\left( \widehat{K}\left\vert \varphi _{0}\right\rangle \right) \right\vert
\right\vert ^{2}}{\left\vert \left\vert T\left( \left\vert \varphi
_{0}\right\rangle \right) \right\vert \right\vert ^{2}+\left\vert \left\vert
T\left( \widehat{K}\left\vert \varphi _{0}\right\rangle \right) \right\vert
\right\vert ^{2}}\;  \label{Tphi0}
\end{equation}

In the present case, we find $\mathcal{T}=0$. Whatever the incident spin,
the tunnel barrier acts as a pure spin rotator, without any spin filter
effect.

The cases of a spin-split quantum well confined between infinite walls and
grown along the $[110]$ direction is discussed in Appendix B2.

\subsubsection{A unified description}

Let us now consider transport in the real conduction band, in Region $I$ or $%
III$. In the case $\gamma =0$, the solution of the Schr\"{o}dinger equation
is $\psi \left( z\right) =\psi ^{\left( 0\right) }\left( z\right)
=a_{j}e^{iq_{0}z}+b_{j}e^{-iq_{0}z}$, $j=1$ or $3$, $b_{3}=0$. When $\gamma $
is non zero, the wave function, in the up-spin channel, has to be of the
form 
\begin{equation}
\psi \left( z\right) =e^{i\vartheta }\left[ a_{j}\left( 1+\alpha \frac{%
\delta q}{q_{0}}\right) e^{iq_{0}z}+b_{j}\left( 1+\beta \frac{\delta q}{q_{0}%
}\right) e^{-iq_{0}z}\right] e^{-i\delta qz}  \label{1110_10}
\end{equation}%
where $e^{i\vartheta }$ is a phase factor. Here again, let us calculate $%
\mathbf{J}$ to the first order in $\delta q$ by making use of Eq. \ref%
{110_g_bis}. Substituting $\psi $ with $\psi ^{(0)}$ in the bracket, one
obtains%
\begin{eqnarray}
\mathbf{J}\left[ \psi \right] &\thickapprox &\mathbf{J}^{f}\left[ \psi %
\right] +\frac{\gamma }{2\hbar }\left[ \left\vert \frac{\partial \psi
^{\left( 0\right) }}{\partial z}\right\vert ^{2}+2q_{0}^{2}\left\vert \psi
^{\left( 0\right) }\right\vert ^{2}\right]  \notag \\
&=&\mathbf{J}^{f}\left[ \psi \right] +2\frac{\gamma _{c}}{\hbar }\delta q%
\left[ 3\left\vert \psi ^{\left( 0\right) }\right\vert ^{2}-2\left(
a_{j}^{\ast }b_{j}e^{2iq_{0}z}+a_{j}b_{j}^{\ast }e^{-2iq_{0}z}\right) \right]
\label{1106_10}
\end{eqnarray}%
\begin{equation}
\mathbf{J}^{f}\left[ \psi \right] =\mathbf{J}^{f}\left[ \psi ^{\left(
0\right) }\right] +4\frac{\gamma _{c}}{\hbar }\delta q\left( \left\vert
a_{j}\right\vert ^{2}\func{Re}\alpha -\left\vert b_{j}\right\vert ^{2}\func{%
Re}\beta \right) -2\frac{\gamma _{c}}{\hbar }\delta q\left\vert \psi
^{\left( 0\right) }\right\vert ^{2}  \label{1106_20}
\end{equation}

For $\func{Re}\alpha =-\func{Re}\beta $, one finds%
\begin{eqnarray}
\mathbf{J}^{f}\left[ \psi \right] &=&\mathbf{J}^{f}\left[ \psi ^{\left(
0\right) }\right] +4\frac{\gamma _{c}}{\hbar }\delta q\func{Re}\alpha \left(
\left\vert a_{j}\right\vert ^{2}+\left\vert b_{j}\right\vert ^{2}\right) -2%
\frac{\gamma _{c}}{\hbar }\delta q\left\vert \psi ^{\left( 0\right)
}\right\vert ^{2}  \notag \\
&=&\mathbf{J}^{f}\left[ \psi ^{\left( 0\right) }\right] +4\frac{\gamma _{c}}{%
\hbar }\delta q\func{Re}\alpha \left[ \left\vert \psi ^{\left( 0\right)
}\right\vert ^{2}-\left( a_{j}^{\ast }b_{j}e^{2iq_{0}z}+a_{j}b_{j}^{\ast
}e^{-2iq_{0}z}\right) \right] -2\frac{\gamma _{c}}{\hbar }\delta q\left\vert
\psi ^{\left( 0\right) }\right\vert ^{2}  \notag \\
&&  \label{110_p}
\end{eqnarray}

Taking $\func{Re}\alpha =-1$ and $\func{Im}\alpha =\func{Im}\beta =0$,%
\begin{equation}
\mathbf{J}\left[ \psi \right] =\mathbf{J}^{f}\left[ \psi ^{\left( 0\right) }%
\right]  \label{110_q}
\end{equation}

In the barrier, we consider (cf Sec. III B 1)%
\begin{equation}
\psi _{B}\left( z\right) =\left[ a_{2}\left( 1-i\frac{Q}{K}\right)
e^{-Kz}+b_{2}\left( 1+i\frac{Q}{K}\right) e^{Kz}\right] e^{iQz}
\label{1106_30}
\end{equation}%
\begin{equation}
\psi _{B}\left( z_{0}\right) e^{-iQz_{0}}=\left(
a_{2}e^{-Kz_{0}}+b_{2}e^{Kz_{0}}\right) -i\frac{Q}{K}\left(
a_{2}e^{-Kz_{0}}-b_{2}e^{Kz_{0}}\right)  \label{1106_40}
\end{equation}%
\begin{equation}
\psi _{B}\left( z_{0}\right) =e^{iQz_{0}}\left[ \psi ^{\left( 0\right)
}\left( z_{0}\right) +i\frac{Q}{K^{2}}\frac{\partial \psi _{B}^{\left(
0\right) }\left( z_{0}\right) }{\partial z}\right] =e^{iQz_{0}}\left[ \psi
^{\left( 0\right) }\left( z_{0}\right) +i\frac{\gamma _{2}}{4\gamma _{2c}}%
\frac{\partial \psi _{B}^{\left( 0\right) }\left( z_{0}\right) }{\partial z}%
\right]  \label{1106_50}
\end{equation}

In the well, let us take (cf Eq. \ref{1110_10}) 
\begin{equation}
\psi _{W}\left( z\right) =e^{iQz_{0}}\left[ a_{j}\left( 1-\frac{\delta q_{j}%
}{q_{j0}}\right) e^{iq_{j0}z}+b_{j}\left( 1+\frac{\delta q_{j}}{q_{j0}}%
\right) e^{-iq_{j0}z}\right] e^{-i\delta q_{j}\left( z-z_{0}\right) }
\label{110_r}
\end{equation}%
where $z_{0}$ is the boundary relevant to the region of the well (e.g. $%
z_{0}=0$ in Region $I$ and $z_{0}=a$ in Region $III$ ). Although we are
still dealing with a unique effective mass and a constant $\gamma $, for the
subsequent discussion, it is convenient to refer to $\gamma $ ($\gamma _{c}$%
) as $\gamma _{2}$ $\left( \gamma _{2c}\right) $ or $\gamma _{j}$ ($\gamma
_{jc}$) where $j=1$ or $3$ in the different regions, and to the wave vectors
as $q_{j0}-\delta q_{j}$ and $-\left( q_{j0}+\delta q_{j}\right) $.%
\begin{equation}
\psi _{W}\left( z_{0}\right) =e^{iQz_{0}}\left[ \left(
a_{j}e^{iq_{j0}z_{0}}+b_{j}e^{-iq_{j0}z_{0}}\right) +\frac{\delta q_{j}}{%
q_{j0}}\left( -a_{j}e^{iq_{j0}z_{0}}+b_{j}e^{-iq_{j0}z_{0}}\right) \right]
\label{1106_60}
\end{equation}%
\begin{equation}
\psi _{W}\left( z_{0}\right) =e^{iQz_{0}}\left[ \psi ^{\left( 0\right)
}\left( z_{0}\right) +i\frac{\delta q_{j}}{q_{j0}^{2}}\frac{\partial \psi
_{W}^{\left( 0\right) }\left( z_{0}\right) }{\partial z}\right] =e^{iQz_{0}}%
\left[ \psi ^{\left( 0\right) }\left( z_{0}\right) +i\frac{\gamma _{j}}{%
4\gamma _{jc}}\frac{\partial \psi _{W}^{\left( 0\right) }\left( z_{0}\right) 
}{\partial z}\right]  \label{1106_70}
\end{equation}

We obtain%
\begin{eqnarray}
\psi _{B}\left( z_{0}\right) -\psi _{W}\left( z_{0}\right) &=&e^{iQz_{0}} 
\left[ \text{ }i\frac{\gamma _{2}}{4\gamma _{2c}}\frac{\partial \psi
_{B}^{\left( 0\right) }\left( z_{0}\right) }{\partial z}-i\frac{\gamma _{j}}{%
4\gamma _{jc}}\frac{\partial \psi _{W}^{\left( 0\right) }\left( z_{0}\right) 
}{\partial z}\right]  \notag \\
&=&e^{iQz_{0}}\frac{i}{4\gamma _{jc}}\frac{\partial \psi _{B}^{\left(
0\right) }\left( z_{0}\right) }{\partial z}\left( \frac{\gamma _{2}\gamma
_{jc}}{\gamma _{2c}}-\frac{\gamma _{j}\gamma _{2c}}{\gamma _{jc}}\right) 
\notag \\
&=&e^{iQz_{0}}\frac{i}{4\gamma _{2c}}\frac{\partial \psi _{W}^{\left(
0\right) }\left( z_{0}\right) }{\partial z}\left( \frac{\gamma _{2}\gamma
_{jc}}{\gamma _{2c}}-\frac{\gamma _{j}\gamma _{2c}}{\gamma _{jc}}\right)
\label{1106_80}
\end{eqnarray}

Here, we have used the relation%
\begin{equation}
\gamma _{2c}\frac{\partial \psi _{B}^{\left( 0\right) }\left( z_{0}\right) }{%
\partial z}=\gamma _{jc}\frac{\partial \psi _{W}^{\left( 0\right) }\left(
z_{0}\right) }{\partial z}  \label{1106_90}
\end{equation}%
which originates from the usual relation expressing current conservation in
the absence of DP field.\cite{Bastard} When $\gamma $ and $\gamma _{c}$
(i.e. $m$) are constant, $\psi _{B}\left( z_{0}\right) -\psi _{W}\left(
z_{0}\right) =0$, which establishes the continuity of the wave function.

Now, let us examine the matching conditions of the derivative%
\begin{equation}
\psi _{B}\left( z\right) =\left[ a_{2}\left( 1-i\frac{Q}{K}\right)
e^{-Kz}+b_{2}\left( 1+i\frac{Q}{K}\right) e^{Kz}\right] e^{iQz}
\label{1106_100}
\end{equation}%
\begin{gather}
\frac{\partial \psi _{B}\left( z_{0}\right) }{\partial z}\qquad \qquad
\qquad \qquad \qquad \qquad \qquad \qquad \qquad \qquad \qquad \qquad \qquad
\qquad \qquad \qquad \qquad  \notag \\
\qquad =e^{iQz_{0}}\left\{ \left[ -Ka_{2}\left( 1-i\frac{Q}{K}\right)
e^{-Kz_{0}}+Kb_{2}\left( 1+i\frac{Q}{K}\right) e^{Kz_{0}}\right] +iQ\left(
a_{2}e^{-Kz_{0}}+b_{2}e^{Kz_{0}}\right) \right\}  \notag \\
=e^{iQz_{0}}\left[ \frac{\partial \psi _{B}^{\left( 0\right) }\left(
z_{0}\right) }{\partial z}+2iQ\psi ^{\left( 0\right) }\left( z_{0}\right) %
\right] \qquad \quad \qquad \qquad \qquad \qquad \qquad \qquad \qquad
\label{1106_110}
\end{gather}%
\begin{eqnarray}
\frac{\partial \psi _{W}\left( z_{0}\right) }{\partial z} &=&e^{iQz_{0}}%
\left\{ iq_{j}\left[ a_{j}\left( 1-\frac{\delta q_{j}}{q_{j0}}\right)
e^{iq_{j0}z_{0}}-b_{j}\left( 1+\frac{\delta q_{j}}{q_{j0}}\right)
e^{-iq_{j0}z_{0}}\right] -i\delta q_{j}\psi ^{\left( 0\right) }\left(
z_{0}\right) \right\}  \notag \\
&=&e^{iQz_{0}}\left[ \frac{\partial \psi _{W}^{\left( 0\right) }\left(
z_{0}\right) }{\partial z}-2i\delta q_{j}\psi ^{\left( 0\right) }\left(
z_{0}\right) \right]  \label{1106_120}
\end{eqnarray}%
\begin{gather}
\gamma _{2c}\frac{\partial \psi _{B}\left( z_{0}\right) }{\partial z}-\gamma
_{jc}\frac{\partial \psi _{W}\left( z_{0}\right) }{\partial z}\qquad \qquad
\qquad \qquad \qquad \qquad \qquad \qquad \qquad \qquad \qquad \qquad \qquad
\notag \\
\qquad =e^{iQz_{0}}\left\{ \left[ \gamma _{2c}\frac{\partial \psi
_{B}^{\left( 0\right) }\left( z_{0}\right) }{\partial z}-\gamma _{jc}\frac{%
\partial \psi _{W}^{\left( 0\right) }\left( z_{0}\right) }{\partial z}\right]
+2i\left( \gamma _{2c}Q+\gamma _{jc}\delta q_{j}\right) \psi ^{\left(
0\right) }\left( z_{0}\right) \right\}  \notag \\
\qquad \qquad =\frac{1}{2}i\left( \gamma _{2}K^{2}+\gamma
_{j}q_{j0}^{2}\right) e^{iQz_{0}}\psi ^{\left( 0\right) }\left( z_{0}\right)
\ \qquad \qquad \qquad \qquad \qquad \qquad \qquad \qquad \qquad
\label{1106_130}
\end{gather}

This is exactly$\ $the jump of the derivative calculated in Eq. \ref{Bord6},
up to the second-order terms. Thus, starting from the standard solution, we
have constructed in a very simple way a wave function which is continuous,
associated to the constant current of probability $\mathbf{J}^{f}\left[ \psi
^{\left( 0\right) }\right] $, and which is the solution to the tunneling
problem.

\subsubsection{Insight into the step-function case}

The case where $\gamma \left( z\right) $ $=\gamma g\left( z\right) $ is not
a constant raises difficult questions. The problem is not to solve Eqs. \ref%
{H_ud_3} but to define a proper Hamiltonian, which has to be Hermitian: This
would not be the case simply by substituting $\gamma $ with $\gamma \left(
z\right) $ in these equations and there are several ways to symmetrize this
Hamiltonian. This is in line with the BDD approach when dealing with an
heterostructure where $m=m(z)$, i.e. where $m$ depends on $z$, for instance $%
m=m_{1}$ in Region $I$ and $m=m_{2\text{ }}$ in Region $II$.\cite%
{Bastard,Balian} In that case, the starting point is the Hamiltonian%
\begin{equation}
\widehat{H}=-\frac{\hbar ^{2}}{2m(z)}\frac{\partial ^{2}}{\partial z^{2}}+V
\label{BDD_100}
\end{equation}

The key idea is to transform this equation by defining the BBD Hamiltonian%
\begin{equation}
\widehat{H}_{BDD}=\frac{\hbar }{2i}\frac{\partial }{\partial z}\widehat{v}+V%
\text{ }  \label{1106_135}
\end{equation}%
where $\widehat{v}$ is defined in Eq. \ref{velo_groupe}. Then, an
integration of the Schr\"{o}dinger equation around the origin, exactly as
performed above, will allow us to show that $\mathbf{J}=\psi ^{\ast }%
\widehat{v}\psi $ is continuous because $\psi $ and $\frac{1}{m\left(
z\right) }\frac{\partial \psi }{\partial z}$ are continuous. The BDD
Hamiltonian guarantees probability-current conservation and the problem
receives sound foundations. Unfortunately, the more complicated form of the
current of probability given in Eq. \ref{110_g} - in particular due to the $%
\left\vert \frac{\partial \psi _{\pm }}{\partial z}\right\vert ^{2}$term -
makes an analogous transformation not obvious so that the general case still
remains an open question. However, let us point out that, when the masses
and the DP-field coefficients are not very different over the three regions
- a frequent situation in heterostructures - through the procedure described
in the preceding subsection, we are able to construct a wave which is
continuous at the boundaries and that conserves the current of probability.
Therefore, this wave is a plausible solution. The principle is first to
solve the envelope-function problem in the absence of DP field, i.e. $%
g\left( z\right) =0$, taking into account the mass discontinuities in
framework of the BDD formalism. This determines the standard function $\psi
^{\left( 0\right) }$ $\left( z\right) $. Secondly, the wave functions in the
different regions are modified according to the rules defined in the
preceding subsection (Eqs. \ref{1106_30} and \ref{110_r}). The current of
probability remains equal to $J^{f}\left[ \psi ^{\left( 0\right) }\right] $
in the three regions. Concerning the continuity of the wave function at the
boundaries, we have at $z_{0}=0$%
\begin{eqnarray}
\frac{\partial \psi _{W}^{\left( 0\right) }\left( 0\right) }{\partial z}
&\thickapprox &2iq_{1}a_{1}  \notag \\
\delta q &\thickapprox &\frac{\gamma _{1}\gamma _{2c}}{\gamma _{2}\gamma
_{1c}}\frac{\gamma _{2}}{4\gamma _{2c}}q_{1}^{2}\quad \text{and}\quad
Q\thickapprox \frac{\gamma _{2}}{4\gamma _{2c}}K^{2}  \label{1106_140}
\end{eqnarray}

Thus (see Eq. \ref{1106_80})%
\begin{equation}
\psi _{B}\left( 0\right) -\psi _{W}\left( 0\right) =-\frac{1}{2}\frac{%
q_{1}a_{1}\text{ }}{\gamma _{2c}}\left( \frac{\gamma _{2}\gamma _{jc}}{%
\gamma _{2c}}-\frac{\gamma _{j}\gamma _{2c}}{\gamma _{jc}}\right)
\label{1106_150}
\end{equation}

We introduce%
\begin{eqnarray}
\gamma _{j,2} &=&\frac{\gamma _{j}+\gamma _{2}}{2}\quad \text{and}\quad
\delta \gamma _{j,2}=\frac{\gamma _{j}-\gamma _{2}}{2}\text{ }  \notag \\
\gamma _{j} &=&\gamma _{j,2}+\delta \gamma _{j,2}\quad \gamma _{2}=\gamma
_{j,2}-\delta \gamma _{j,2}  \label{1106_160}
\end{eqnarray}%
\begin{eqnarray}
\Gamma _{j,2} &=&\frac{\gamma _{jc}+\gamma _{2c}}{2}\quad \text{and}\quad
\delta \Gamma _{j,2}=\frac{\gamma _{jc}-\gamma _{2c}}{2}\text{ }  \notag \\
\gamma _{jc} &=&\Gamma _{j,2}+\delta \Gamma _{j,2}\quad \gamma _{2c}=\Gamma
_{j,2}-\delta \Gamma _{j,2}  \label{1106_170}
\end{eqnarray}%
\begin{eqnarray}
\psi _{B}\left( 0\right) -\psi _{W}\left( 0\right) &=&-\frac{1}{2}\frac{%
q_{1}a_{1}\text{ }}{\gamma _{2c}}\gamma _{1,2}\left[ \frac{\left( 1-\frac{%
\delta \gamma _{1,2}}{\gamma _{1,2}}\right) \left( 1+\frac{\delta \Gamma
_{1,2}}{\Gamma _{1,2}}\right) }{1-\frac{\mathbf{\delta \Gamma }_{1,2}}{%
\Gamma _{1,2}}}-\frac{\left( 1+\frac{\delta \gamma _{1,2}}{\gamma _{1,2}}%
\right) \left( 1-\frac{\delta \Gamma _{1,2}}{\Gamma _{1,2}}\right) }{1+\frac{%
\mathbf{\delta \Gamma }_{1,2}}{\Gamma _{1,2}}}\right]  \notag \\
&=&\frac{q_{1}a_{1}\text{ }}{\gamma _{2c}}\gamma _{1,2}\left( \frac{\delta
\gamma _{1,2}}{\gamma _{1,2}}-2\frac{\delta \Gamma _{1,2}}{\Gamma _{1,2}}%
\right)  \notag \\
&\thickapprox &\frac{q_{1}a_{1}\text{ }}{\gamma _{2c}}\gamma _{2}\left( 
\frac{\delta \gamma _{1,2}}{\gamma _{1,2}}-2\frac{\delta \Gamma _{1,2}}{%
\Gamma _{1,2}}\right) =4a_{1}\frac{q_{1}}{K}\frac{Q}{K}\left( \frac{\delta
\gamma _{1,2}}{\gamma _{1,2}}-2\frac{\delta \Gamma _{1,2}}{\Gamma _{1,2}}%
\right)  \label{1106_180}
\end{eqnarray}

At $z_{0}=a$, the situation is similar with%
\begin{equation}
\frac{\partial \psi _{W}^{\left( 0\right) }\left( a\right) }{\partial z}%
=iq_{3}a_{3}e^{iQa}e^{iq_{3}a}  \label{1106_190}
\end{equation}

In the case where $\frac{\delta \gamma _{j,2}}{\gamma _{j,2}}$ and $\frac{%
\delta \Gamma _{j,2}}{\Gamma _{j,2}}$ are small and considered as
first-order terms, the discontinuities are third-order terms which can be
safely neglected.

\subsubsection{Quasi-classical picture (Regions $I$ and $III$ without
sizable spin splitting)}

In the case where Regions $I$ and $III$ have no sizable spin splitting, we
develop a quasi-classical picture of the tunneling process. For an up spin $%
Q=Q_{\uparrow }$, so that the wave function in the barrier writes as%
\begin{eqnarray}
\psi _{II\,+}(z) &=&\left[ a_{2}\left( 1-\frac{iQ}{K}\right)
e^{-Kz}+b_{2}\left( 1+\frac{iQ}{K}\right) e^{Kz}\right] e^{iQz}  \notag \\
&=&\psi _{II}^{\left( 0\right) }(z)e^{iQz}+\left( \frac{iQ}{K}\frac{1}{K}%
\frac{\partial }{\partial z}\psi _{II}^{\left( 0\right) }(z)\right) e^{iQz}
\label{solu_up}
\end{eqnarray}

The wave function for the down spin is obtained by replacing $Q$ by $-Q$. We
can combine the two spin channels to build the quasi-classical solution $%
\psi ^{c}\left( z\right) $ corresponding to an incident wave with a spin
lying in the plane perpendicular to the DP field%
\begin{equation}
\psi _{I}^{c}\left( z\right) =\left( \lambda \uparrow +\lambda ^{\ast
}\downarrow \right) \psi _{I}^{\left( 0\right) }\left( z\right) =S_{\lambda
}(0)\psi _{I}^{\left( 0\right) }\left( z\right)  \label{inci_plan}
\end{equation}%
which yields%
\begin{equation}
\psi _{II}^{c}(z)=\psi _{II}^{\left( 0\right) }(z)S_{\lambda }(z)-\frac{Q}{K}%
\frac{1}{K}\frac{\partial }{\partial z}\psi _{II}^{\left( 0\right) }(z)i\hat{%
K}S_{\lambda }(z)  \label{solu_plan}
\end{equation}%
Defining%
\begin{equation}
\tan \theta =\frac{Q}{K}\thickapprox \theta  \label{tan_theta}
\end{equation}%
we can write to the first order%
\begin{equation}
\psi _{II}^{c}(z)=\cos \theta \psi _{II}^{\left( 0\right) }(z)S_{\lambda
}(z)-\sin \theta \frac{1}{K}\frac{\partial }{\partial z}\psi _{II}^{\left(
0\right) }(z)i\hat{K}S_{\lambda }\left( z\right)  \label{solu_plan1}
\end{equation}%
The transmitted wave is%
\begin{equation}
\psi _{III}^{c}=\left( \lambda e^{iQa}\uparrow +\lambda ^{\ast
}e^{-iQa}\downarrow \right) a_{3}e^{iqz}=S_{\lambda }(a)\psi _{III}^{\left(
0\right) }  \label{trans_plan}
\end{equation}%
The incident wave corresponds to a spin lying in the $\Pi _{\mathbf{\chi }}$
plane, normal to\textbf{\ }$\chi \left( \mathbf{e}_{110}\right) $. An
important result is that the transmitted wave has the spin $S_{\alpha }(a)$,
i.e. rotated by the angle $-2Qa$. We can estimate the angle $2Qa$ $\approx
0.2\left( K/1\text{ \AA\ }^{-1}\right) ^{2}\left( a/1\ \text{\AA }\right) $
in GaAs along the $[110]$ direction, the largest reasonable value of $K$
being smaller than $0.1$ \AA $^{-1}$, a value beyond which the
spin-splitting in $k^{3}$ is no longer valid.\textbf{\ }\textit{The
spin-split barrier appears to exert a spin torque which produces a rotation
of the spin of the transmitted electron around the quantization axis, which
is the direction of the DP field.} \textit{There is no spin transmission
asymmetry }. The spin-orbit-split barrier acts as a spin rotator inside the $%
\mathbf{\Pi }_{\mathbf{\chi }}$ plane.\textbf{\ }This has some analogy with
the reflection of a neutron beam on a ferromagnetic mirror discussed in Ref. 
%TCIMACRO{\TeXButton{\onlinecite{CTDL-ferro}}{\onlinecite{CTDL-ferro}} }%
%BeginExpansion
\onlinecite{CTDL-ferro}
%EndExpansion
which physically results from spin precession during the time spent by the
evanescent wave inside the barrier. But, in this example, this
straightforwardly arises from the difference in the reflection and
transmission coefficients for the two spin eigenstates. Anyway, this spin
precession provides an estimation of the tunnel time $\tau $, by using this
built-in Larmor clock.\cite{Buttiker} The effective field is determined
through: $\hbar \Omega \approx 2\gamma \left\vert \overline{\chi }%
\right\vert $ whereas $\Omega \tau =2Qa\thickapprox \left\vert a\gamma 
\overline{\chi }_{e}/\gamma _{c}\right\vert K^{2}$. We find $\tau \approx
\left\vert a\hbar /2\gamma _{c}\right\vert \left\vert \overline{\chi }_{e}/%
\overline{\chi }\right\vert K^{2}$. In the $\left[ 110\right] $ direction, $%
\overline{\chi }_{e}=1/2$ (see Sec. \ref{energie}) so that $\tau \approx
\left\vert a\hbar /4\gamma _{c}K\right\vert \approx 10^{-18}\left( a/1\,%
\text{\AA }\right) \left( 1\,\text{\AA }^{-1}/K\right) $ s.

We recognize that the in-plane solution belongs to the subspace of
free-electron-current conserving waves studied in Sec. \ref{J_subspace}. In
that sense, we have restored a "classical" tunneling process. Note that $%
\mathbf{J=J}^{f}$ is a constant, but the classical "magnetic current" in
Region $II$ , $\delta \mathbf{J}^{f}(z)=\mathbf{J}_{\uparrow }^{f}(z)-%
\mathbf{J}_{\downarrow }^{f}(z)$ is not and undergoes a discontinuity at the
boundaries. Quite generally for any two-component spinor $\psi =\psi
_{+}\uparrow +\psi _{-}\downarrow $ with $\psi _{+}=\Phi \,e^{iQz}$ and $%
\psi _{-}=\Phi \,e^{-iQz}$ 
\begin{subequations}
\label{F_400}
\begin{align}
\mathbf{J}^{f}\left[ \psi _{\pm }\right] & =\frac{\hbar }{m}\func{Im}\left[
\psi _{\pm }^{\ast }\nabla \psi _{\pm }\right]  \label{F_400_a} \\
\delta \mathbf{J}^{f}& =\mathbf{J}^{f}\left[ \psi _{+}\right] -\mathbf{J}^{f}%
\left[ \psi _{-}\right] =\frac{\hbar }{m}\func{Im}\left[ \psi _{+}^{\ast
}\nabla \psi _{+}-\psi _{-}^{\ast }\nabla \psi _{-}\right]  \label{F_400_b}
\\
\delta \mathbf{J}^{f}& =\frac{\hbar }{m}\func{Im}\left[ \left( \psi
_{+}\uparrow -\psi _{-}\downarrow \right) ^{\dagger }\nabla \left( \psi
_{+}\uparrow +\psi _{-}\downarrow \right) \right] =\frac{1}{m}\func{Re}\left[
\psi ^{\dagger }\left( \widehat{\mathbf{p}}\cdot \widehat{\mathbf{\sigma }}%
\right) \psi \right]  \label{F_400_c}
\end{align}

Thus, the jump of $\delta \mathbf{J}^{f}$ is 
\end{subequations}
\begin{equation}
\left[ \delta \mathbf{J}^{f}\right] _{z_{0}^{W}}^{z_{0}^{B}}=\frac{\hbar }{m}%
\func{Im}\left\{ \left( \widehat{\sigma }_{z}\psi \right) _{z_{0}}^{\dagger }%
\left[ \nabla \psi \right] _{z_{0}^{W}}^{z_{0}^{B}}\right\} =\frac{1}{m}%
\func{Re}\left[ \psi ^{\dagger }\left( \widehat{\mathbf{p}}\cdot \widehat{%
\mathbf{\sigma }}\right) \psi \right] _{z_{0}^{W}}^{z_{0}^{B}}  \label{F_420}
\end{equation}%
More explicitly, we have $\delta \mathbf{J}_{I}^{f}=\delta \mathbf{J}%
_{III}^{f}=0,\ \delta \mathbf{J}_{II}^{f}(z)\simeq 2\left\vert \lambda
\right\vert ^{2}\dfrac{\hbar Q}{m}\left\vert \dfrac{1}{K}\dfrac{\partial
\Phi _{II}(z)}{\partial z}\right\vert ^{2}$. This can be viewed as a
kinetic-momentum transfer along the internal-field direction during the
tunnel process, in strong analogy with the spin transfer resulting from spin
torque in ferromagnetic structures, as introduced by Slonczewski and Berger.%
\cite{Slonczewski, Berger}.

\section{Ortho process: [001]-oriented barrier under almost normal incidence 
\label{ortho}}

It is not possible to stay in simple band schemes - like in Fig. 3 - as $%
\mathbf{\xi }$ has to be conserved: The relevant scheme is drawn in Fig. 6.
To simplify without altering the physics of interest, the component of the
wave vector normal to [001] is taken parallel to [100]. The spin is
quantized along the $Oz$ axis, taken parallel to [001]. As shown below, the
eigenstates of the spin are in a direction normal to $Oz$. The energy writes%
\begin{equation}
E=-\gamma _{c}(\mathcal{K}^{2}-\xi ^{2})\pm \gamma \xi \mathcal{K}\sqrt{%
\mathcal{K}^{2}-\xi ^{2}}  \label{O_100}
\end{equation}%
\begin{equation}
\left[ E+\gamma _{c}(\mathcal{K}^{2}-\xi ^{2})\right] ^{2}=\left( \gamma \xi 
\mathcal{K}\right) ^{2}\left( \mathcal{K}^{2}-\xi ^{2}\right)  \label{O_120}
\end{equation}%
where the generic wave vector is $\xi \mathbf{e}_{100}+i\mathcal{K}\mathbf{e}%
_{001}$. This equation may admit 4 real roots, $\pm K$, and $\pm K^{\prime }$%
. The states of the four wave vectors $(\xi ,0,iK)\uparrow _{\mathbf{k}}$, $%
(\xi ,0,-iK)\uparrow _{\mathbf{k}^{\ast }}$, $(\xi ,0,iK^{\prime
})\downarrow _{\mathbf{k}^{\prime }}$, $(\xi ,0,-iK^{\prime })\downarrow _{%
\mathbf{k}^{\prime \ast }}\ $have the same energies: $K$ and $K^{\prime }$
are such that $E\uparrow (K)=E\downarrow (K^{\prime })$. Note that Kramers
conjugate states, which would involve $-\xi $, are not relevant because $\xi 
$ is conserved. We introduce $K_{0}=(K^{\prime }+K)/2$ and $\delta
K=K^{\prime }-K$ (note that this definition differs by a factor of 2 of the
definition used in Sec. III, where $2\delta q=q^{\prime }-q$ ; The choice
made in the present section makes more easy the comparison with the results
derived in Ref. 
%TCIMACRO{\TeXButton{\onlinecite{Perel03}}{\onlinecite{Perel03}}}%
%BeginExpansion
\onlinecite{Perel03}%
%EndExpansion
). We assume that $K^{\prime }>K>0$ so that $\delta K>0$. Moreover, as in
Ref. 
%TCIMACRO{\TeXButton{\onlinecite{Perel03}}{\onlinecite{Perel03}}}%
%BeginExpansion
\onlinecite{Perel03}%
%EndExpansion
, the incident-wave energy is smaller than half of the barrier energy, which
means that $q<K$.

As recognized by Perel' et al., the tunneling problem admits simple $C^{1}$
solutions under the approximation $\xi /K_{0}\ll 1$. Besides, the spin
asymmetry which originates from the spin-orbit interaction is characterized
by the ratio $\delta K/K_{0}$, which, from band-structure calculations\cite%
{Jancu} and from spin-precession experiments,\cite{Jusserand,Riechert} is
known to be small, i.e. $\delta K/K_{0}\ll 1$. We further assume that $%
aK_{0}\ $is not small compared to unity, which corresponds to a barrier of
small transparency, and consequently we have $\exp \left( -2a\,K_{0}\right)
\ll $ $1$. These three quantities, $\xi /K_{0}$, $\delta K/K_{0}$, and $\exp
\left( -2a\,K_{0}\right) $ will be hereafter taken as first-order quantities
and we will look for solutions to the first order only. This does not imply
that the quantity $a\,\delta K=(a\,K_{0})(\delta K/K_{0})$, which is of
crucial interest as it characterizes the spin selectivity of the barrier (as
illustrated by the simple evaluation indicated below), is smaller than
unity. In the physical problem, we consider electron tunneling under
off-normal incidence and the angle of incidence is significant only when $q$
and $\xi $ are of the same order, which means $q/K_{0}\ll 1$. We shall use
this additional approximation only when it will be necessary to get
analytical expressions of the wave vectors (Subsec. \ref{first}).
Intuitively, if we start with an unpolarized electron beam, the up- (down-)
spin electrons merge from the barrier with an amplitude of probability
almost proportional to $\exp -a\,K$ (respectively $\exp -aK^{\prime }$) so
that the current asymmetry - which, in this case, is also the polarization $%
\Pi $ of the current - is given by%
\begin{equation}
\Pi \approx \frac{e^{-2aK}-e^{-2aK^{\prime }}}{e^{-2aK}+e^{-2aK\prime }}%
=\tanh a\delta K  \label{PI_100}
\end{equation}

Indeed, in Ref. 
%TCIMACRO{\TeXButton{\onlinecite{Perel03}}{\onlinecite{Perel03}}}%
%BeginExpansion
\onlinecite{Perel03}%
%EndExpansion
, it is found that the polarization $\mathcal{P}$ of the transmitted
current, when the primary beam is not polarized, is $\mathcal{P}\thickapprox
\tanh a\,\delta K$ (see below Eq. \ref{Perel0}). In practical cases, $%
a\,\delta K$ cannot be larger than a (often small) fraction of unity.
Nevertheless, in the calculation, we do not put any restrictive assumption
on $a\,\delta K$ (which is not assumed to be a first-order quantity) and we
will calculate eigenvectors, when required, as a power expansion in $%
a\,\delta K$, but, obviously, we keep in mind that the first-order term will
generally be sufficient to reach a reasonable accuracy.

\subsection{Zeroth-order wave functions\label{zero}}

The wave vectors $K$ and $K^{\prime }$ are related through the equation $%
\left( K^{\prime }>K\text{ and assuming }\gamma >0\text{ for the sake of
simplicity}\right) $%
\begin{equation}
-\gamma _{c}(K^{2}-\xi ^{2})-\gamma \xi K\sqrt{K^{2}-\xi ^{2}}=-\gamma
_{c}(K^{\prime 2}-\xi ^{2})+\gamma \xi K^{\prime }\sqrt{K^{\prime \,2}-\xi
^{2}}  \label{001-80}
\end{equation}%
or%
\begin{equation}
\gamma _{c}(K^{\prime \,2}-K^{2})=\gamma \xi \left( K\sqrt{K^{2}-\xi ^{2}}%
+K^{\prime }\sqrt{K^{\prime \,2}-\xi ^{2}}\right)  \label{001-120}
\end{equation}

Up to the first order in $\delta K/K_{0}$, Eq. \ref{001-120} writes as%
\begin{equation}
2\gamma _{c}K_{0}\delta K=2\gamma \xi K_{0}\sqrt{\left( K_{0}^{2}-\xi
^{2}\right) }  \label{001-130}
\end{equation}%
or$\hspace{-1cm}$%
\begin{equation}
\delta K=\dfrac{\gamma \xi K_{0}}{\gamma _{c}}\sqrt{\left( 1-\dfrac{\xi ^{2}%
}{K^{2}}\right) }\thickapprox \dfrac{\gamma \xi K_{0}}{\gamma _{c}}
\label{001-141}
\end{equation}

We now calculate the eigenvectors. Let us write

$k=(\xi ,0,\eta i\mathcal{K})$\ with $\eta =\pm 1$, $\mathcal{K}=K$ or $%
K^{\prime }$, $\xi ,\,K$ and $K^{\prime }>0$. $\mathbf{\chi }=\mathcal{K}\xi
(K,\,0,\,i\eta \xi )$.

The eigenvalues of $\widehat{\mathbf{\sigma }}\cdot \mathbf{\chi }=%
\begin{bmatrix}
\chi _{z} & \chi _{x}-i\chi _{y} \\ 
\chi _{x}+i\chi _{y} & -\chi _{z}%
\end{bmatrix}%
$ are $\pm \xi \mathcal{K}\sqrt{\mathcal{K}^{2}-\xi ^{2}}$.

To the first order in $\xi /K_{0}$, the normalized eigenvectors $%
c_{1}\uparrow +c_{2}\downarrow =%
\begin{bmatrix}
c_{1} \\ 
c_{2}%
\end{bmatrix}%
$ are such that

\begin{equation}
\begin{array}{ccccccccc}
\begin{array}{l}
\text{wave} \\ 
\text{vector}%
\end{array}
&  & \left[ 
\begin{array}{c}
\xi \\ 
0 \\ 
iK%
\end{array}%
\right] & \quad & \left[ 
\begin{array}{c}
\xi \\ 
0 \\ 
-iK%
\end{array}%
\right] & \quad & \left[ 
\begin{array}{c}
\xi \\ 
0 \\ 
iK^{\prime }%
\end{array}%
\right] & \quad & \left[ 
\begin{array}{c}
\xi \\ 
0 \\ 
-iK^{\prime }%
\end{array}%
\right] \\ 
&  &  &  &  &  &  &  &  \\ 
\text{spin}\times \sqrt{2} &  & 
\begin{bmatrix}
1+\dfrac{i\xi }{2K} \\ 
1-\dfrac{i\xi }{2K}%
\end{bmatrix}
&  & 
\begin{bmatrix}
1-\dfrac{i\xi }{2K} \\ 
1+\dfrac{i\xi }{2K}%
\end{bmatrix}
&  & 
\begin{bmatrix}
1-\dfrac{i\xi }{2K^{\prime }} \\ 
-\left( 1+\dfrac{i\xi }{2K^{\prime }}\right)%
\end{bmatrix}
&  & 
\begin{bmatrix}
1+\dfrac{i\xi }{2K^{\prime }} \\ 
-\left( 1-\dfrac{i\xi }{2K^{\prime }}\right)%
\end{bmatrix}%
\end{array}
\label{001-100}
\end{equation}

Observe that $\uparrow _{\mathbf{k}}$ and $\downarrow _{\mathbf{k}}$ are not
orthogonal (even in a first-order calculation - compare the first term to
the third one after substituting $K^{\prime }$ with $K$). Inside the barrier
the wave function is of the shape $\Psi ^{II}(\mathbf{r})=e^{i\mathbf{\xi
\cdot \rho }}\,\Psi ^{II}(z)$ and%
\begin{align}
\Psi ^{II}(z)& =A_{2}%
\begin{bmatrix}
1+\dfrac{i\xi }{2K} \\ 
1-\dfrac{i\xi }{2K}%
\end{bmatrix}%
e^{-Kz}+B_{2}%
\begin{bmatrix}
1-\dfrac{i\xi }{2K} \\ 
1+\dfrac{i\xi }{2K}%
\end{bmatrix}%
e^{Kz}  \notag \\
& \qquad \qquad +\tilde{A}_{2}%
\begin{bmatrix}
1-\dfrac{i\xi }{2K^{\prime }} \\ 
-(1+\dfrac{i\xi }{2K^{\prime }})%
\end{bmatrix}%
e^{-K^{\prime }z}+\tilde{B}_{2}%
\begin{bmatrix}
1+\dfrac{i\xi }{2K^{\prime }} \\ 
-(1-\dfrac{i\xi }{2K^{\prime }})%
\end{bmatrix}%
e^{K^{\prime }z}  \notag \\
& =\left[ A_{2}e^{-Kz}+B_{2}e^{Kz}+\frac{i\xi }{2K^{\prime }}\left( -\tilde{A%
}_{2}e^{-K^{\prime }z}+\tilde{B}_{2}e^{K^{\prime }z}\right) \right] 
\begin{bmatrix}
1 \\ 
1%
\end{bmatrix}
\notag \\
& \qquad \qquad +\left[ \frac{i\xi }{2K}\left(
A_{2}e^{-Kz}-B_{2}e^{Kz}\right) +\tilde{A}_{2}e^{-K^{\prime }z}+\tilde{B}%
_{2}e^{K^{\prime }z}\right] 
\begin{bmatrix}
1 \\ 
-1%
\end{bmatrix}
\label{OO_100}
\end{align}

Outside the barrier, we are looking for the solution of the shape:%
\begin{equation}
\Psi ^{I}(z)=A_{1}%
\begin{bmatrix}
1 \\ 
1%
\end{bmatrix}%
e^{iqz}+B_{1}%
\begin{bmatrix}
1 \\ 
1%
\end{bmatrix}%
e^{-iqz}+\tilde{A}_{1}%
\begin{bmatrix}
1 \\ 
-1%
\end{bmatrix}%
e^{iqz}+\tilde{B}_{1}%
\begin{bmatrix}
1 \\ 
-1%
\end{bmatrix}%
e^{-iqz}  \label{OO_120}
\end{equation}%
and%
\begin{equation}
\Psi ^{III}(z)=A_{3}%
\begin{bmatrix}
1 \\ 
1%
\end{bmatrix}%
e^{iqz}+\tilde{A}_{3}%
\begin{bmatrix}
1 \\ 
-1%
\end{bmatrix}%
e^{iqz}  \label{OO_140}
\end{equation}

The wave function writes as 
\begin{subequations}
\label{001_100}
\begin{eqnarray}
\Psi ^{I}(z) &=&\left[ A_{1}e^{iqz}+B_{1}e^{-iqz}\right] \left[ 1\;1\right]
^{t}+\left[ \tilde{A}_{1}e^{iqz}+\tilde{B}_{1}e^{-iqz}\right] \left[ 1\;-1%
\right] ^{t}  \label{001_100_a} \\
\Psi ^{II}(z) &=&\left[ A_{2}e^{-Kz}+B_{2}e^{Kz}+\frac{i\xi }{2K^{\prime }}%
\left( -\tilde{A}_{2}e^{-K^{\prime }z}+\tilde{B}_{2}e^{K^{\prime }z}\right) %
\right] \left[ 1\;1\right] ^{t}  \notag \\
&&\qquad +\left[ \frac{i\xi }{2K}\left( A_{2}e^{-Kz}-B_{2}e^{Kz}\right) +%
\tilde{A}_{2}e^{-K^{\prime }z}+\tilde{B}_{2}e^{K^{\prime }z}\right] \left[
1\;-1\right] ^{t}  \label{001_100_b} \\
\Psi ^{III}(z) &=&\left[ A_{3}e^{iqz}\right] \left[ 1\;1\right] ^{t}+\left[ 
\tilde{A}_{3}e^{iqz}\right] \left[ 1\;-1\right] ^{t}  \label{001_100_c}
\end{eqnarray}

The continuity of the wave function (Eqs. \ref{001_100}) and of its
derivative at $z=0$ and $z=a$ provides a linear system of 8 equations. A
full discussion is given in Appendix \ref{Coeff_0} . This calculation has
strong similarities with Slonczewski's\cite{Slonczewski} approach of the
tunneling between two ferromagnets separated by a barrier, because we deal
with two coupled spin channels.

\subsection{Polarization}

The transmission asymmetry $\mathcal{T}$ is 
\end{subequations}
\begin{equation}
\mathcal{T}=\dfrac{\left\vert t^{+}\right\vert ^{2}-\left\vert
t^{-}\right\vert ^{2}}{\left\vert t^{+}\right\vert ^{2}+\left\vert
t^{-}\right\vert ^{2}}  \label{polar}
\end{equation}%
with $\left\vert t^{+}\right\vert ^{2}=\left\vert A_{3}^{+}\right\vert ^{2}$
(calculated when $A_{1}=1,$ $\widetilde{A}_{1}=0$) and $\left\vert
t^{-}\right\vert ^{2}=\left\vert \widetilde{A_{3}^{-}}\right\vert ^{2}$%
(calculated when $A_{1}=0,$ $\widetilde{A}_{1}=1$). All the coefficients $%
A_{j}$ and $\widetilde{A}_{j}$ are calculated in Appendix \ref{Coeff_0}.

To the zeroth order in $\xi /K_{0}$, $t^{\pm }=t_{0}^{\pm }$, $\mathcal{T=T}%
_{0}$. Now%
\begin{equation}
\left\vert t_{0}^{+}\right\vert ^{2}=\left\vert \frac{4qKe^{-Ka}}{\left(
K-iq\right) ^{2}}\right\vert ^{2}\;;\;\left\vert t_{0}^{-}\right\vert
^{2}=\left\vert \frac{4qK^{\prime }e^{-K^{\prime }a}}{\left( K^{\prime
}-iq\right) ^{2}}\right\vert ^{2}  \label{t0-2}
\end{equation}%
and we get the result of Ref. 
%TCIMACRO{\TeXButton{\onlinecite{Perel03}}{\onlinecite{Perel03}} }%
%BeginExpansion
\onlinecite{Perel03}
%EndExpansion
namely%
\begin{equation}
\mathcal{T}_{0}=\tanh a\,\delta K  \label{Perel0}
\end{equation}

Up to the first order in $\xi /K_{0}$ , $t^{\pm }=t_{1}^{\pm }$, $\mathcal{%
T=T}_{1}$, $\left\vert t_{1}^{\pm }\right\vert ^{2}=\left( \left\vert
A_{3}^{\pm }\right\vert ^{2}+\left\vert \widetilde{A}_{3}^{\pm }\right\vert
^{2}\right) $ but $\left\vert \widetilde{A}_{3}^{+}\right\vert ^{2}$ and $%
\left\vert A_{3}^{-}\right\vert ^{2}$ are of second order in $\xi /K_{0}$ so
that, \textit{up to the first order in} $\xi /K_{0}\,$, \textit{the result
is the same as for the zeroth order}: $\mathcal{T}_{0}=\mathcal{T}_{1}$.

It is easy to show that this transmission asymmetry is nothing but the spin
polarization of the transmitted beam when the primary beam is unpolarized, $%
\mathcal{T}_{0}=\mathcal{T}_{1}=\mathcal{P}$. As we have only assumed that $%
q<K_{0}$, we may wonder why the ratio $q/K_{0}$ does not appear in $\mathcal{%
P}$. The answer is given if we perform the calculation one order further in $%
\delta K/K_{0}\ll 1$. Then, a lengthy calculation leads to%
\begin{equation}
\mathcal{P}=\dfrac{\tanh a\,\delta K+\frac{K_{0}-q}{K_{0}+q}\frac{\delta K}{%
K_{0}}}{1+\frac{K_{0}-q}{K_{0}+q}\frac{\delta K}{K_{0}}\tanh a\,\delta K}
\label{PerelHH}
\end{equation}

In the limit where $\delta K/K_{0}$ is negligible, $\mathcal{P}=\tanh
a\,\delta K$ is recovered.

Let us consider the transmission of a primary electron beam with an initial
current polarization $P_{i}$ through a spin-filtering structure
characterized by the transmission coefficients $e^{-2aK^{\prime }}$ $\left(
e^{-2aK}\right) $ for up (down) - spin electrons. As the incident up-
(down-) spin current is proportional to $1+\mathcal{P}_{i}$ (respectively $1-%
\mathcal{P}_{i}$), the current polarization of the emerging beam is simply
given by $\mathcal{P}$%
\begin{equation}
\mathcal{P}=\frac{(1+\mathcal{P}_{i})e^{-2aK}-(1-\mathcal{P}%
_{i})e^{-2aK^{\prime }}}{(1+\mathcal{P}_{i})e^{-2aK}+(1-\mathcal{P}%
_{i})e^{-2aK^{\prime }}}=\frac{\mathcal{P}_{i}+\Pi }{1+\Pi \,\mathcal{P}_{i}}
\label{PI_140}
\end{equation}%
where $\Pi $ is given by Eq. \ref{PI_100}. The above formula yielding the
polarization of the transmitted beam is a standard expression for spin
filters (in spin polarimetry, $\Pi $ is referred to as the Sherman function).%
\cite{Drouhin} Thus, $\mathcal{P}$ in Eq. \ref{PerelHH} appears to result
from the combination of a primary-electron-beam polarization\textbf{\ }$%
\mathcal{P}_{i}\approx -\delta K/K_{0}$\ when $q/K_{0}\ll 1$, which does not
depend on the barrier thickness, with the spin asymmetry of the material, $%
\Pi =\tanh \left( a\,\delta K\right) $.\textbf{\ }The initial polarization $%
-\delta K/K_{0}$ could be straightforwardly understood as resulting from the
band mismatch, an interface effect. If this analogy provides us with a
useful physical insight, it must, however, be realized that the above
calculation is only valid when $\exp aK_{0}\gg 1$ and cannot be extrapolated
to $a=0$. In any case, it is clear that $P_{i}$ builds up in the early stage
of the transport process.

\subsection{$\protect\xi /K_{0}$ first-order wave function\label{first}}

It is shown in Appendix \ref{Coeff_0} that there is no $\xi /K_{0}$
first-order term in $A_{2}$, $A_{3}$, $B_{1}$, and $B_{2}$. We are therefore
going to calculate $\xi /K_{0}$ first-order terms in $\widetilde{B}_{1}$, $%
\widetilde{A}_{2}$, $\widetilde{B}_{2}$, and $\widetilde{A}_{3}$. To be
consistent with Subsec. \ref{zero}, we assume that $A_{1}\neq 0$ and $%
\widetilde{A}_{1}=0$. We obviously have to invert the role of $K$ and $%
K^{\prime }$ if we start from $A_{1}=0$ and $\widetilde{A}_{1}\neq 0$. Let
us recall that the calculation is performed with $\delta K/K_{0}\ll 1$ which
is always true and $\xi /K_{0}\ll 1$\textbf{.}

Eqs. \ref{Perel_100.5}, \ref{Perel_100.6}, \ref{Perel_100.7}, \ref%
{Perel_100.8} give 
\begin{subequations}
\label{Perel_120}
\begin{eqnarray}
\hspace{-2cm}-\widetilde{A}_{2}\left( 1-\dfrac{iq}{K^{\prime }}\right) +%
\widetilde{B}_{2}\left( 1+\dfrac{iq}{K^{\prime }}\right) &=&\dfrac{i\xi }{%
2K^{\prime }}\left( 1-\dfrac{iq}{K}\right) A_{2}+\dfrac{i\xi }{2K^{\prime }}%
\left( 1+\dfrac{iq}{K}\right) B_{2}  \notag \\
&&  \label{Perel_120.1} \\
-\widetilde{A}_{2}\left( 1+\dfrac{iq}{K^{\prime }}\right) e^{-K^{\prime }a}+%
\widetilde{B}_{2}\left( 1-\dfrac{iq}{K^{\prime }}\right) e^{K^{\prime }a} &=&%
\dfrac{i\xi }{2K^{\prime }}\left( 1+\dfrac{iq}{K}\right) A_{2}e^{-Ka}+\dfrac{%
i\xi }{2K^{\prime }}\left( 1-\dfrac{iq}{K}\right) B_{2}e^{Ka}  \notag \\
&&  \label{Perel_120.2}
\end{eqnarray}

The determinant of the system defined by Eqs. \ref{Perel_120} is 
\end{subequations}
\begin{equation}
Det=\left( 1+\dfrac{iq}{K^{\prime }}\right) ^{2}e^{-K^{\prime }a}-\left( 1-%
\dfrac{iq}{K^{\prime }}\right) ^{2}e^{K^{\prime }a}  \label{O_140}
\end{equation}%
which differs from zero, therefore $\widetilde{A}_{2}$ and $\widetilde{B}%
_{2} $ can be calculated.

We assume $a\neq 0$ (the case $a=0$ has no interest) and we obtain 
\begin{equation}
\widetilde{A}_{2}=-\dfrac{i\xi }{2K^{\prime }}\left[ A_{2}e^{a\,\delta K/2}%
\dfrac{\sinh \left( K_{0}a\right) }{\sinh \left( K^{\prime }a\right) }%
+B_{2}e^{K_{0}a}\dfrac{\sinh \left( a\,\delta K/2\right) }{\sinh \left(
K^{\prime }a\right) }\right]  \label{O_180_a}
\end{equation}

and%
\begin{equation}
\widetilde{B}_{2}=\dfrac{i\xi }{2K^{\prime }}\left[ A_{2}e^{-K_{0}a}\dfrac{%
\sinh \left( a\,\delta K/2\right) }{\sinh \left( K^{\prime }a\right) }%
+B_{2}e^{-a\,\delta K/2}\dfrac{\sinh \left( K_{0}a\right) }{\sinh \left(
K^{\prime }a\right) }\right]  \label{O_180_b}
\end{equation}

Noticing that i) $\xi /K^{\prime }=\xi /K_{0}\left( 1+\delta K/2K\right) 
\QTR{sl}{\approx }\left( \xi /K_{0}\right) \left( 1-\delta K/2K_{0}\right)
\approx \xi /K_{0}$ (the same result holds for $\xi /K\approx \xi /K_{0}$),
ii) $a\,\delta K\ll a\,K$, iii) $A_{2}\varpropto A_{1}$ (Eq. \ref{ab_a2}),
and iv) $B_{2}\varpropto A_{1}\exp \left( -2Ka\right) $ (Eq. \ref{ab_b2}),
we get 
\begin{subequations}
\label{O_200}
\begin{align}
\widetilde{A}_{2}& \approx -\dfrac{i\xi }{2K_{0}}A_{2}e^{a\,\delta K/2}%
\dfrac{\sinh \left( K_{0}a\right) }{\sinh \left( K^{\prime }a\right) }
\label{O_200a} \\
\widetilde{B}_{2}& =\dfrac{i\xi }{2K_{0}}\left[ A_{2}e^{-K_{0}a}\dfrac{\sinh
\left( a\,\delta K/2\right) }{\sinh \left( K^{\prime }a\right) }%
+B_{2}e^{-a\,\delta K/2}\dfrac{\sinh \left( K_{0}a\right) }{\sinh \left(
K^{\prime }a\right) }\right]  \label{O_200b}
\end{align}

From now on we assume that $\exp K_{0}a\gg 1$ so that $\sinh K_{0}a/\sinh
K^{\prime }a=\exp -a\,\delta K/2$ and 
\end{subequations}
\begin{equation}
\widetilde{A}_{2}\approx -\dfrac{i\xi }{2K_{0}}A_{2}  \label{Perel_200}
\end{equation}

A lengthy$\ $calculation shows that:

i) $\widetilde{B}_{1}$ is proportional to $\left( \xi /K_{0}\right) \left(
\delta K/K_{0}\right) $ and therefore is negligible. However, we can note
that $\widetilde{B}_{1}$ is not strictly equal to zero so that the reflected
wave has a $\left[ 1\;-1\right] ^{t}$ component even though the incident
wave has only a $\left[ 1\;1\right] ^{t}$ component.

ii)%
\begin{equation}
\widetilde{B}_{2}\approx \frac{i\xi }{2K_{0}}e^{-a\delta K}\left[ 2\frac{iK+q%
}{iK-q}e^{-a\,\delta K/2}\sinh \frac{a\,\delta K}{2}+1\right] B_{2}
\label{Perel_220}
\end{equation}

We furthermore assume that $q/K_{0}\ll 1$ so that%
\begin{equation}
\widetilde{B}_{2}\approx \frac{i\xi }{2K_{0}}e^{-a\,\delta K}\left[
2-e^{-a\,\delta K}\right] B_{2}  \label{Perel_240}
\end{equation}%
and eventually%
\begin{equation}
\widetilde{A}_{3}=\dfrac{i\xi }{2K_{0}}\left( \sinh \frac{a\,\delta K}{2}%
-2\sinh ^{2}\frac{a\,\delta K}{2}\right) A_{3}  \label{Perel_260}
\end{equation}

There is no assumption on $a\delta K$ in Eq. \ref{Perel_260}.

We note that, as $\widetilde{A}_{3}$ differs from zero, the incident wave
with only a $\left[ 1\;1\right] ^{t}$ spin component is transmitted with a
component along the $\left[ 1\;-1\right] ^{t}$ spin direction. This means
there is not a pure spin-filter effect along the $x$-quantization axis.\cite%
{filter}

\section{Conclusion\label{Co}}

Electron tunneling in a semiconductor with no inversion symmetry and in the
presence of spin-orbit coupling involves complex wave vectors in the
barrier. In directions where the D'yakonov-Perel' (DP) field is non zero,
the problem becomes highly non-trivial. We have distinguished two particular
types of tunnel processes: Para-type process where we have one-dimensional
tunneling with a complex wave vector and ortho-type process associated with
a complex wave vector with orthogonal real and imaginary components. For a
para-process, the DP field is a complex vector but it remains collinear to a
real direction so that the eigenvectors are orthogonal spin states. We have
shown that, along the $[110]$ direction no $C^{1}$ solution exists. The
expression of the current of probability is re-examined, proper boundary
conditions are derived, and a treatment of heterostructures is proposed.
Quasi-classical states are shown to be in-plane solutions, which imply a
pure spin rotation of the transmitted beam around the direction of the DP
field. In the $\left[ 110\right] $ direction, there is no spin-filter
effect. This contrasts with the situation in the real conduction band where
the spin-splitting is maximum along $[110]$. For an ortho-process, the DP
field is a complex vector, which is not collinear to any real direction, and
the eigenvectors of the Hamiltonian are no longer orthogonal spin states.
Moreover, the evanescent eigenvectors are not associated with the same spin
depending whether they propagate from left to right or from right to left.
In this case, we have derived a first-order solution to the tunnel problem,
which has strong similarities with standard off-normal tunneling, and an
almost pure spin-filter effect was demonstrated, a conclusion consistent
with the result of Perel' et al.\cite{Perel03} whose expression for the
transmitted polarization has been corrected by the introduction of an
initial interface polarization.

All these questions should now be addressed experimentally and we think that
experiments are within reach. For instance, further developments of the
study of the polarization of a reflected spin-polarized electron beam can be
considered, in line with the\textbf{\ }measurements reported in Ref. 
%TCIMACRO{\TeXButton{\onlinecite{Joly}}{\onlinecite{Joly}}}%
%BeginExpansion
\onlinecite{Joly}%
%EndExpansion
. Polarized-luminescence experiments in quantum wells grown along the $[110]$
axis could also bring valuable information, as well as measurements on
resonant-tunneling devices or photogavalnic-effect measurements in coupled
quantum wells.\cite{Katsnelson,Diehl,GLA,ROM} The results derived in the
present article provide new insight in spin-dependent tunneling in solids
whereas they also open stimulating perspectives for spin manipulation in
tunnel devices.

%TCIMACRO{\TeXButton{\begin{acknowledgements}}{\begin{acknowledgements}}}%
%BeginExpansion
\begin{acknowledgements}%
%EndExpansion

HJD is a member of the D\'{e}l\'{e}gation G\'{e}n\'{e}rale pour l'Armement.
We are deeply indebted to Michel Dyakonov for an illuminating discussion. We
thank Catherine Bouton-Drouhin, Henri Jaffr\`{e}s, and Andr\'{e} Roug\'{e}
for useful advice, and Jean-No\"{e}l Chazalviel and Travis Wade for a
careful reading of the manuscript.

%TCIMACRO{\TeXButton{\end{acknowledgements}}{\end{acknowledgements}}}%
%BeginExpansion
\end{acknowledgements}%
%EndExpansion

%TCIMACRO{\TeXButton{\appendix}{\appendix}}%
%BeginExpansion
\appendix%
%EndExpansion

\section{Evanescent band in the $\left[ 110\right] $ direction\label{110_dir}%
}

Let us write $\mathbf{k}=(Q+iK)\mathbf{e}$, having in mind $\mathbf{e}$
along the $[110]$ direction: $\mathbf{e=e}_{110}=\frac{1}{\sqrt{2}}\left[ 110%
\right] $. We have to find the relation between $Q$ and $K$ to get a real
eigenvalue of the Hamiltonian $\widehat{H}$. This real eigenvalue is the
energy. The Hamiltonian $\widehat{H}$\ writes as%
\begin{equation}
\widehat{H}=\gamma _{c}\left( Q+iK\right) ^{2}+\gamma \,\widehat{\mathbf{%
\sigma }}\mathbf{\cdot \chi }=\gamma _{c}\left( Q+iK\right) ^{2}+\gamma 
\overline{\chi }_{\mathbf{e}}\left( Q+iK\right) ^{3}\widehat{\mathbf{\sigma }%
}\mathbf{\cdot e}_{\mathbf{\chi }}  \label{Q_100}
\end{equation}

$\mathbf{e}_{\mathbf{\chi }}=\mathbf{\chi }/\left\vert \left\vert \mathbf{%
\chi }\right\vert \right\vert $ (provided $\left\vert \left\vert \mathbf{%
\chi }\right\vert \right\vert \neq 0$). $\overline{\chi }_{e}$ , a
dimensionless parameter, depends on the direction.\ If $\mathbf{e=e}_{110}$, 
$\mathbf{\chi }$ is parallel to $\mathbf{e}_{1\overline{1}0}$ with $%
\overline{\chi }_{\mathbf{e}}=1/2$.

The eigenvalues are%
\begin{equation}
\mathcal{E}\left( \mathbf{k}\right) =\gamma _{c}\left( Q+iK\right)
^{2}+\epsilon \overline{\chi }_{\mathbf{e}}\gamma \left( Q+iK\right) ^{3}
\label{Q_120}
\end{equation}

The spin is quantized along $\gamma \mathbf{e}_{\mathbf{\chi }}$\ so that $%
\epsilon \gamma >0$\ corresponds to the spin $\uparrow $\ and $\epsilon
\gamma <0$\ corresponds to the\ spin $\downarrow $. Separating the real and
imaginary parts of the eigenvalue, we obtain%
\begin{equation}
\func{Re}\mathcal{E}\left( \mathbf{k}\right) =\gamma
_{c}(Q^{2}-K^{2})+\epsilon \overline{\chi }_{\mathbf{e}}\gamma
(Q^{3}-3QK^{2})  \label{Q_140}
\end{equation}%
\begin{equation}
\func{Im}\mathcal{E}\left( \mathbf{k}\right) =2\gamma _{c}QK+\epsilon 
\overline{\chi }_{\mathbf{e}}\gamma (3Q^{2}K-K^{3})  \label{Q_160}
\end{equation}

Looking for the real-energy lines, we have the equation%
\begin{equation}
\func{Im}\mathcal{E}\left( \mathbf{k}\right) =0\quad \Longrightarrow \quad
2\gamma _{c}Q+\epsilon \overline{\chi }_{\mathbf{e}}\gamma (3Q^{2}-K^{2})=0
\label{Q_180}
\end{equation}%
\begin{equation}
K^{2}=3Q^{2}+2\epsilon \frac{\gamma _{c}}{\gamma \overline{\chi }_{\mathbf{e}%
}}Q\quad \left( =3Q^{2}+\epsilon \,4\frac{\gamma _{c}}{\gamma }Q\text{ if }%
\mathbf{e=e}_{110}\right)  \label{Q_200}
\end{equation}

Eq. \ref{Q_200} is the relation between $Q$ and $K$ we were looking for. The
energy is%
\begin{align}
\hspace{-1cm}E_{\epsilon }(Q)& =-\epsilon 8\overline{\mathbf{\chi }}%
_{e}\gamma Q^{3}-8\gamma _{c}Q^{2}-\epsilon 2\dfrac{\gamma _{c}^{2}}{\gamma 
\overline{\chi }_{\mathbf{e}}}Q  \notag \\
& =-\epsilon 4\gamma Q^{3}-8\gamma _{c}Q^{2}-\epsilon 4\dfrac{\gamma _{c}^{2}%
}{\gamma }Q\text{ if }\mathbf{e=e}_{110}  \label{Q_205}
\end{align}

For a given $E(Q)$ value, we have \textit{two} possible choices of $K$,%
\begin{equation}
K=\pm \sqrt{3Q^{2}+2\epsilon \frac{\gamma _{c}}{\gamma \overline{\chi }_{%
\mathbf{e}}}Q}\;\left( =\pm \sqrt{3Q^{2}+\epsilon \,4\frac{\gamma _{c}}{%
\gamma }Q}\text{ if }\mathbf{e=e}_{110}\right)  \label{Q_240}
\end{equation}

Let us note that $\left\vert \epsilon \,4\left( \gamma /\gamma _{c}\right)
Q\right\vert \gg 3Q^{2}$ so that $\left\vert Q\right\vert \ll \left\vert
K\right\vert $ and%
\begin{equation}
K\approx \pm \sqrt{\left( 4\epsilon \gamma _{c}/\gamma \right) Q}
\label{Q_250}
\end{equation}

The sign of $\epsilon \,\gamma $ determines the sign of $Q$ $\left( \gamma
_{c}>0\right) $.\ As stated above $\epsilon \gamma >0$, which corresponds to
spin $\uparrow $, gives $Q>0$ whereas $\epsilon \gamma <0$, which
corresponds to spin $\downarrow $, gives $Q<0$.

We have the symmetry property 
\begin{equation}
E_{\pm }(Q)=E_{\mp }(-Q)  \label{Q_260}
\end{equation}

The study of the function $E(Q)$ is straightforward and we take $\epsilon
=-1 $ in the following, the other case being deduced by symmetry:%
\begin{equation}
\frac{\text{d}E_{-}(Q)}{\text{d}Q}=24\overline{\mathbf{\chi }}_{e}\gamma
Q^{2}-16\gamma _{c}Q+2\dfrac{\gamma _{c}^{2}}{\gamma \overline{\chi }_{%
\mathbf{e}}}\quad \left( =12\gamma Q^{2}-16\gamma _{c}Q+4\dfrac{\gamma
_{c}^{2}}{\gamma }\text{ if }\mathbf{e=e}_{110}\right)  \label{Q_280}
\end{equation}

The roots $Q_{1}$ and $Q_{2}$ of the derivative are%
\begin{eqnarray}
Q_{1} &=&\frac{\gamma _{c}}{2\gamma \overline{\chi }_{\mathbf{e}}}\;\text{,}%
\;Q_{2}=\frac{\gamma _{c}}{6\gamma \overline{\chi }_{\mathbf{e}}}  \notag \\
\quad Q_{1} &=&\frac{\gamma _{c}}{\gamma }\;\text{,}\;Q_{2}=\frac{\gamma _{c}%
}{3\gamma }\text{\ if }\mathbf{e=e}_{110}  \label{Q_305}
\end{eqnarray}

Incidentally we note that%
\begin{equation}
E_{-}(Q_{1})=0  \label{Q_320}
\end{equation}

The corresponding curve is plotted in Fig. 4. It must be realized that we
are only dealing with evanescent states, which correspond to a negative
energy. Thus, for a given energy $E<0$, we have two possible $Q$-values, $%
\pm Q$, each associated with a given spin subband.

Finally, we find that, at a given energy, we have \textit{exactly} the four
possible states, the wave vectors of which are $(Q\pm iK)$ with spin $%
\uparrow $ and $(-Q\pm iK)$ with spin $\downarrow $, the latter being
obtained from the former through $\widehat{K}$. In short%
\begin{equation}
E_{\uparrow }\left( \mathbf{k}\right) =E_{\uparrow }\left( \mathbf{k}^{\ast
}\right) =E_{\downarrow }\left( -\mathbf{k}\right) =E_{\downarrow }\left( -%
\mathbf{k}^{\ast }\right)  \label{A14}
\end{equation}

\section{Continuity equation and definition of the probability current\label%
{defi_curr}}

\subsection{Definition of the probability current}

Consider a Hamiltonian given by%
\begin{equation}
\hat{H}=\sum\limits_{j}a_{j}\hat{p}_{j}+\sum\limits_{j,k}b_{jk}\hat{p}_{j}%
\hat{p}_{k}+\sum\limits_{j,k,l}c_{jkl}\hat{p}_{j}\hat{p}_{k}\hat{p}_{l}+V
\label{hamitonian_gen}
\end{equation}%
where $a_{j}$, $b_{jk}$, $c_{jkl}$ are Hermitian matrices, invariant under
permutation of the indices $i$, $j$, $k$ $,$ and where $V$ is real. We
define the velocity operator%
\begin{equation}
\hat{v}_{j}=\frac{\partial \hat{H}}{\partial p_{j}}=a_{j}+2\sum%
\limits_{k}b_{jk}\hat{p}_{k}+3\sum\limits_{k,l}c_{jkl}\hat{p}_{k}\hat{p}_{l}
\label{veclo_oper}
\end{equation}

It will be useful to take the following notations:%
\begin{eqnarray}
\left\vert \psi \right) &=&\psi _{1}\left( \mathbf{r}\right) \uparrow +\psi
_{2}\left( \mathbf{r}\right) \downarrow =\left[ 
\begin{array}{c}
\psi _{1}\left( \mathbf{r}\right) \\ 
\psi _{2}\left( \mathbf{r}\right)%
\end{array}%
\right] ,\ \left\vert \phi \right) =\phi _{1}\left( \mathbf{r}\right)
\uparrow +\phi _{2}\left( \mathbf{r}\right) \downarrow =\left[ 
\begin{array}{c}
\phi _{1}\left( \mathbf{r}\right) \\ 
\phi _{2}\left( \mathbf{r}\right)%
\end{array}%
\right]  \notag \\
\left( \psi \right\vert &=&\left[ \psi _{1}^{\ast }\left( \mathbf{r}\right)
\;\;\psi _{2}^{\ast }\left( \mathbf{r}\right) \right] ,\quad \left( \phi
\right\vert =\left[ \phi _{1}^{\ast }\left( \mathbf{r}\right) \;\;\phi
_{2}^{\ast }\left( \mathbf{r}\right) \right] ,\ \left( \phi |\psi \right)
=\phi _{1}^{\ast }\left( \mathbf{r}\right) \psi _{1}\left( \mathbf{r}\right)
+\phi _{2}^{\ast }\left( \mathbf{r}\right) \psi _{2}\left( \mathbf{r}\right)
\notag \\
\left( \psi |\psi \right) &=&\psi _{1}^{\ast }\left( \mathbf{r}\right) \psi
_{1}\left( \mathbf{r}\right) +\psi _{2}^{\ast }\left( \mathbf{r}\right) \psi
_{2}\left( \mathbf{r}\right) =\left\vert \psi \left( \mathbf{r}\right)
\right\vert ^{2}=\left\vert \psi \right\vert ^{2}  \notag \\
\left\vert \widehat{p}\psi \right) &=&\left[ 
\begin{array}{c}
\widehat{p}\psi _{1}\left( \mathbf{r}\right) \\ 
\widehat{p}\psi _{2}\left( \mathbf{r}\right)%
\end{array}%
\right] ,\ \left( \widehat{p}\psi \right\vert =\left[ \widehat{p}^{\ast
}\phi _{1}^{\ast }\left( \mathbf{r}\right) \quad \widehat{p}^{\ast }\phi
_{2}^{\ast }\left( \mathbf{r}\right) \right] =\left[ -\widehat{p}\phi
_{1}^{\ast }\left( \mathbf{r}\right) \quad -\widehat{p}\phi _{2}^{\ast
}\left( \mathbf{r}\right) \right]  \notag \\
\left( \widehat{p}\phi |\widehat{p}\psi \right) &=&\left[ \widehat{p}^{\ast
}\phi _{1}^{\ast }\left( \mathbf{r}\right) \right] \left[ \widehat{p}\psi
_{1}\left( \mathbf{r}\right) \right] +\left[ \widehat{p}^{\ast }\phi
_{2}^{\ast }\left( \mathbf{r}\right) \right] \left[ \widehat{p}\psi
_{2}\left( \mathbf{r}\right) \right]  \notag \\
&=&\left[ -\widehat{p}\phi _{1}^{\ast }\left( \mathbf{r}\right) \right] %
\left[ \widehat{p}\psi _{1}\left( \mathbf{r}\right) \right] +\left[ -%
\widehat{p}\phi _{2}^{\ast }\left( \mathbf{r}\right) \right] \left[ \widehat{%
p}\psi _{2}\left( \mathbf{r}\right) \right]  \label{pseudobk}
\end{eqnarray}

The Schr\"{o}dinger equation is%
\begin{eqnarray}
i\hbar \frac{\partial \left\vert \psi \right) }{\partial t}
&=&\sum\limits_{j}a_{j}\hat{p}_{j}\left\vert \psi \right)
+\sum\limits_{j,k}b_{jk}\hat{p}_{j}\hat{p}_{k}\left\vert \psi \right)
+\sum\limits_{j,k,l}c_{jkl}\hat{p}_{j}\hat{p}_{k}\hat{p}_{l}\left\vert \psi
\right)  \notag \\
-i\hbar \frac{\partial \left( \psi \right\vert }{\partial t}
&=&\sum\limits_{j}\left( \widehat{p}_{j}\psi \right\vert
a_{j}+\sum\limits_{j,k}\left( \hat{p}_{j}\hat{p}_{k}\psi \right\vert
b_{jk}+\sum\limits_{j,k,l}\left( \hat{p}_{j}\hat{p}_{k}\hat{p}_{l}\psi
\right\vert c_{jkl}  \label{Schro_gen}
\end{eqnarray}

The continuity equation can be written%
\begin{eqnarray}
i\hbar \left[ \left( \psi |\frac{\partial }{\partial t}\psi \right) +\left( 
\frac{\partial }{\partial t}\psi |\psi \right) \right] &=&i\hbar \frac{%
\partial \left\vert \psi \right\vert ^{2}}{\partial t}  \notag \\
&=&\sum\limits_{j}\left[ \left( \psi |a_{j}\hat{p}_{j}\psi \right) -\left( 
\hat{p}_{j}\psi |a_{j}\psi \right) \right]  \notag \\
&&+\sum\limits_{j,k}\left[ \left( \psi |b_{jk}\hat{p}_{j}\hat{p}_{k}\psi
\right) -\left( \hat{p}_{j}\hat{p}_{k}\psi |b_{jk}\psi \right) \right] 
\notag \\
&&+\sum\limits_{j,k,l}\left[ \left( \psi |c_{jkl}\hat{p}_{j}\hat{p}_{k}\hat{p%
}_{l}\psi \right) -\left( \hat{p}_{j}\hat{p}_{k}\hat{p}_{l}\psi |b_{jk}\psi
\right) \right]  \label{continu_gen}
\end{eqnarray}

Note that\ 
\begin{equation}
\left( \psi |a_{j}\hat{p}_{j}\psi \right) =\left( a_{j}\hat{p}_{j}\psi |\psi
\right) ^{\ast }=\left( \hat{p}_{j}\psi |a_{j}\psi \right) ^{\ast }
\label{rela_1}
\end{equation}%
or%
\begin{equation}
\left( \hat{p}_{j}\psi |a_{j}\psi \right) =\left( \psi |a_{j}\hat{p}_{j}\psi
\right) ^{\ast }  \label{rela_2}
\end{equation}

Similarly%
\begin{equation}
\left( \hat{p}_{j}\hat{p}_{k}\psi |b_{jk}\psi \right) =\left( \psi |b_{jk}%
\hat{p}_{j}\hat{p}_{k}\psi \right) ^{\ast }\quad \text{;}\quad \left( \hat{p}%
_{j}\hat{p}_{k}\hat{p}_{l}\psi |b_{jk}\psi \right) =\left( \psi |c_{jkl}\hat{%
p}_{j}\hat{p}_{k}\hat{p}_{l}\psi \right) ^{\ast }  \label{rela_3}
\end{equation}

Therefore%
\begin{equation}
\frac{\partial \left\vert \psi \right\vert ^{2}}{\partial t}=\frac{2}{\hbar }%
\func{Im}\left[ \sum\limits_{j}\left( \psi |a_{j}\hat{p}_{j}\psi \right)
+\sum\limits_{j,k}\left( \psi |b_{jk}\hat{p}_{j}\hat{p}_{k}\psi \right)
+\sum\limits_{j,k,l}\left( \psi |c_{jkl}\hat{p}_{j}\hat{p}_{k}\hat{p}%
_{l}\psi \right) \right]  \label{diver_J_1}
\end{equation}

The probability current $\mathbf{J}$ has to satisfy%
\begin{eqnarray}
\mathbf{\nabla \cdot J} &=&-\frac{2}{\hbar }\func{Im}\left[
\sum\limits_{j}\left( \psi |a_{j}\hat{p}_{j}\psi \right)
+\sum\limits_{j,k}\left( \psi |b_{jk}\hat{p}_{j}\hat{p}_{k}\psi \right)
+\sum\limits_{j,k,l}\left( \psi |c_{jkl}\hat{p}_{j}\hat{p}_{k}\hat{p}%
_{l}\psi \right) \right]  \notag \\
&=&\mathbf{\nabla \cdot J}^{\left( 1\right) }+\mathbf{\nabla \cdot J}%
^{\left( 2\right) }+\mathbf{\nabla \cdot J}^{\left( 3\right) }
\label{diver_J_2}
\end{eqnarray}

From the expression of the velocity operator, we tentatively define the
probability current as%
\begin{equation}
\mathbf{\tilde{J}}_{j}=\left[ \frac{1}{2}\left( \psi |a_{j}\psi \right)
+\sum\limits_{k}\left( \psi |b_{jk}\hat{p}_{k}\psi \right) +\frac{3}{2}%
\sum\limits_{j,k,l}\left( \hat{p}_{k}\psi |c_{jkl}\hat{p}_{k}\psi \right) %
\right] +cc  \label{J_essai}
\end{equation}%
where $cc$ refers to the complex conjugate. We calculate%
\begin{equation}
\mathbf{\nabla \cdot \tilde{J}}=\sum\limits_{j}\nabla _{j}\mathbf{\tilde{J}}%
_{j}=\frac{i}{\hbar }\sum\limits_{j}\hat{p}_{j}\mathbf{\tilde{J}}_{j}
\label{diver_gen}
\end{equation}

Let us consider the first term%
\begin{equation}
\mathbf{\tilde{J}}_{j}^{\left( 1\right) }=\frac{1}{2}\left( \psi |a_{j}\psi
\right) +cc=\left( \psi |a_{j}\psi \right)  \label{J_essai_first}
\end{equation}%
\begin{eqnarray}
\sum\limits_{j}\nabla _{j}\mathbf{\tilde{J}}_{j}^{\left( 1\right) } &=&\frac{%
i}{\hbar }\sum\limits_{j}\hat{p}_{j}\left( \psi |a_{j}\psi \right) =\frac{i}{%
\hbar }\sum\limits_{j}\left( \psi |a_{j}\hat{p}_{j}\psi \right) -\left( \hat{%
p}_{j}\psi |a_{j}\psi \right)  \notag \\
&=&-\frac{2}{\hbar }\func{Im}\sum\limits_{j}\left( \psi |a_{j}\hat{p}%
_{j}\psi \right) =\mathbf{\nabla \cdot J}^{\left( 1\right) }
\label{diverr_Jes_first}
\end{eqnarray}

The second term gives%
\begin{equation}
\mathbf{\tilde{J}}_{j}^{\left( 2\right) }=\sum\limits_{k}\left( \psi |b_{jk}%
\hat{p}_{k}\psi \right) +cc=\sum\limits_{k}\left[ \left( \psi |b_{jk}\hat{p}%
_{k}\psi \right) +\left( \hat{p}_{k}\psi |b_{jk}\psi \right) \right]
\label{J_essaie_second}
\end{equation}%
\begin{eqnarray}
\sum\limits_{j}\nabla _{j}\mathbf{\tilde{J}}_{j}^{\left( 2\right) } &=&\frac{%
i}{\hbar }\sum\limits_{j,k}\left[ \left( \psi |b_{jk}\hat{p}_{j}\hat{p}%
_{k}\psi \right) -\left( \hat{p}_{j}\psi |b_{jk}\hat{p}_{k}\psi \right)
+\left( \hat{p}_{k}\psi |b_{jk}\hat{p}_{j}\psi \right) -\left( \hat{p}_{j}%
\hat{p}_{k}\psi |b_{jk}\psi \right) \right]  \notag \\
&=&-\frac{2}{\hbar }\func{Im}\sum\limits_{j}\left( \psi |b_{jk}\hat{p}_{j}%
\hat{p}_{k}\psi \right) =\mathbf{\nabla \cdot J}^{\left( 2\right) }
\label{diverr_Jes_second}
\end{eqnarray}

Concerning to the third term%
\begin{equation}
\mathbf{\tilde{J}}_{j}^{\left( 3\right) }=\frac{3}{2}\sum\limits_{k,l}\left(
p_{k}\psi |c_{jkl}\hat{p}_{l}\psi \right) +cc=3\sum\limits_{k,l}\left(
p_{k}\psi |c_{jkl}\hat{p}_{l}\psi \right)  \label{J_essaie_third}
\end{equation}%
\begin{equation}
\sum\limits_{j}\nabla _{j}\mathbf{\tilde{J}}_{j}^{\left( 3\right) }=\frac{3i%
}{\hbar }\sum\limits_{k,l}\left[ \left( \hat{p}_{k}\psi |c_{jkl}\hat{p}_{j}%
\hat{p}_{l}\psi \right) -\left( \hat{p}_{j}\hat{p}_{k}\psi |c_{jkl}\hat{p}%
_{l}\psi \right) \right] \neq \mathbf{\nabla \cdot J}^{\left( 3\right) }
\label{diverr_Jes_third}
\end{equation}

Let us now consider the quantity%
\begin{eqnarray}
\sum\limits_{jkl}\hat{p}_{j}\hat{p}_{k}\hat{p}_{l}\left( \psi |c_{jkl}\psi
\right) &=&\sum\limits_{jkl}\left[ \left( \psi |c_{jkl}\hat{p}_{j}\hat{p}_{k}%
\hat{p}_{l}\psi \right) -\left( \hat{p}_{j}\hat{p}_{k}\hat{p}_{l}\psi
|c_{jkl}\psi \right) \right]  \notag \\
&&\qquad -3\sum\limits_{jkl}\left[ \left( \hat{p}_{j}\psi |c_{jkl}\hat{p}_{k}%
\hat{p}_{l}\psi \right) -\left( \hat{p}_{j}\hat{p}_{k}\psi |c_{jkl}\hat{p}%
_{l}\psi \right) \right]  \notag \\
&=&\sum\limits_{j}\hat{p}_{j}\sum\limits_{kl}\hat{p}_{k}\hat{p}_{l}\left(
\psi |c_{jkl}\psi \right)  \notag \\
&=&\frac{\hbar }{i}\sum\limits_{j}\nabla _{j}\sum\limits_{kl}\hat{p}_{k}\hat{%
p}_{l}\left( \psi |c_{jkl}\psi \right)  \label{Add_quant_1}
\end{eqnarray}

We have%
\begin{equation}
\sum\limits_{j}\nabla _{j}\left[ \sum\limits_{k,l}\hat{p}_{k}\hat{p}%
_{l}\left( \psi |c_{jkl}\psi \right) +\mathbf{\tilde{J}}_{j}^{\left(
3\right) }\right] =-\frac{2}{\hbar }\func{Im}\sum\limits_{j,k,l}\left( \psi
|c_{jkl}\hat{p}_{j}\hat{p}_{k}\hat{p}_{l}\psi \right) =\mathbf{\nabla \cdot J%
}^{\left( 3\right) }  \label{diver_Jvrai_third}
\end{equation}

Thus, we can define%
\begin{equation}
\mathbf{J}_{j}^{\left( 3\right) }=\mathbf{\tilde{J}}_{j}^{\left( 3\right)
}+\sum\limits_{k,l}\hat{p}_{k}\hat{p}_{l}\left( \psi |c_{jkl}\psi \right)
\label{J_vrai_third}
\end{equation}

Finally, the $j$ component of the probability current can be taken as%
\begin{equation}
\mathbf{J}_{j}=\left[ \frac{1}{2}\left( \psi |a_{j}\psi \right)
+\sum\limits_{k}\left( \psi |b_{jk}\hat{p}_{k}\psi \right) +\frac{3}{2}%
\sum\limits_{j,k,l}\left( \hat{p}_{k}\psi |c_{jkl}\hat{p}_{k}\psi \right) +%
\frac{1}{2}\sum\limits_{k,l}\hat{p}_{k}\hat{p}_{l}\left( \psi |c_{jkl}\psi
\right) \right] +cc  \label{J_vrai_1}
\end{equation}%
or%
\begin{equation}
\mathbf{J}_{j}=\mathbf{J}_{j}^{f}+\left( \psi |a_{j}\psi \right)
+3\sum\limits_{j,k,l}\left( \hat{p}_{k}\psi |c_{jkl}\hat{p}_{k}\psi \right)
+\sum\limits_{k,l}\hat{p}_{k}\hat{p}_{l}\left( \psi |c_{jkl}\psi \right)
\label{J_vrai_2}
\end{equation}

\subsection{Quantum well grown in the $[110]$ direction\label{IQW110}}

To illustrate some simple consequences, we apply the preceding results to
the practical case of quantum wells grown in the $[110]$ direction. First,
let us point out that, in this case, a direct calculation of the current of
probability is straightforward%
\begin{eqnarray}
i\hbar \frac{\partial \psi _{\pm }}{\partial t} &=&-\gamma _{c}\frac{%
\partial ^{2}\psi _{\pm }}{\partial z^{2}}\pm \frac{1}{2}i\gamma \frac{%
\partial ^{3}\psi _{\pm }}{\partial z^{3}}  \notag \\
-i\hbar \frac{\partial \psi _{\pm }^{\ast }}{\partial t} &=&-\gamma _{c}%
\frac{\partial ^{2}\psi ^{\pm }}{\partial z^{2}}\mp \frac{1}{2}i\gamma \frac{%
\partial ^{3}\psi _{\pm }^{\ast }}{\partial z^{3}}  \label{Schro}
\end{eqnarray}

Multiplying the first equation by $\psi _{\pm }^{\ast }$, the second
equation by $\psi _{\pm }$ and subtracting them, we obtain%
\begin{equation}
i\hbar \left( \psi _{\pm }^{\ast }\frac{\partial \psi _{\pm }}{\partial t}%
+\psi _{\pm }\frac{\partial \psi _{\pm }^{\ast }}{\partial t}\right)
=-\gamma _{c}\left( \psi _{\pm }^{\ast }\frac{\partial ^{2}\psi _{\pm }}{%
\partial z^{2}}-\psi _{\pm }\frac{\partial ^{2}\psi _{\pm }^{\ast }}{%
\partial z^{2}}\right) \pm \frac{1}{2}i\gamma \left( \psi _{\pm }^{\ast }%
\frac{\partial ^{3}\psi _{\pm }}{\partial z^{3}}+\psi _{\pm }\frac{\partial
^{3}\psi _{\pm }^{\ast }}{\partial z^{3}}\right)  \label{conser_pro_1}
\end{equation}

or%
\begin{eqnarray}
-\mathbf{\nabla \cdot J}_{\pm } &=&\frac{\partial \left\vert \psi
\right\vert ^{2}}{\partial t}=-\frac{\gamma _{c}}{i\hbar }\left( \psi _{\pm
}^{\ast }\frac{\partial ^{2}\psi _{\pm }}{\partial z^{2}}-\psi _{\pm }\frac{%
\partial ^{2}\psi _{\pm }^{\ast }}{\partial z^{2}}\right) +\frac{1}{2}\frac{%
\gamma }{\hbar }\left( \psi _{\pm }^{\ast }\frac{\partial ^{3}\psi _{\pm }}{%
\partial z^{3}}+\psi _{\pm }\frac{\partial ^{3}\psi _{\pm }^{\ast }}{%
\partial z^{3}}\right)  \notag \\
&=&-\mathbf{\nabla \cdot }\left[ \mathbf{J}_{\pm }^{f}\pm \frac{\gamma }{%
2\hbar }\left( 3\left\vert \frac{\partial }{\partial z}\psi _{\pm
}\right\vert ^{2}-\frac{\partial ^{2}}{\partial z^{2}}\left\vert \psi _{\pm
}\right\vert ^{2}\right) \right]  \label{conser_pro_2p}
\end{eqnarray}

We consider a well made of a spin-split semiconductor (GaAs) confined
between infinite walls located at $z=0$ and $z=a$. At energy $E$, for a
given spin, the wave function $\phi \left( z\right) $ consists of a
combination of eigenstates associated to the wave vectors $q\left( E\right) $
and $-q^{\prime }\left( E\right) $ (see Fig. 6, upper part) which satisfy%
\begin{equation}
\gamma _{c}q^{2}+\frac{1}{2}\gamma q^{3}=\gamma _{c}q^{\prime 2}-\frac{1}{2}%
\gamma q^{\prime 3}  \label{Ener_well}
\end{equation}

The wave function writes%
\begin{equation}
\phi (z)=A\,e^{iqz}+B\,e^{-iq^{\prime }z}  \label{fonc_well_1}
\end{equation}

and verifies the boundary condition: $\phi (0)=\phi (a)=0$ so that $A=-B$
and $q+q^{\prime }=n\frac{2\pi }{a}$, or%
\begin{equation}
\phi (z)=2iA\;\sin \left( \frac{n\pi }{a}z\right) e^{-i\delta q\,z}
\label{fonc_well_2}
\end{equation}

A straightforward calculation gives%
\begin{equation}
-\mathbf{\nabla \cdot J=}\frac{\partial \left\vert \phi \right\vert ^{2}}{%
\partial t}=-\frac{2}{\hbar }\left\vert A\right\vert ^{2}\sin \left\{
(q+q^{\prime })z\right\} \left[ \gamma _{c}\left( q^{2}-q^{\prime 2}\right) +%
\frac{1}{2}\gamma \left( q^{\prime 3}+q^{3}\right) \right] =0  \label{PC_60}
\end{equation}%
due to the energy expression (Eq. \ref{Ener_well}). The probability current $%
\mathbf{J}$ is conserved as it should. However, a calculation of $\mathbf{J}$
according to Eq. \ref{conser_pro_2p} yields 
\begin{equation}
\mathbf{J}_{\pm }\mathbf{=}\pm \frac{1}{2\hbar }\mathbf{\gamma }\left\vert 
\mathbf{A}\right\vert ^{2}\left( q+q^{\prime }\right) ^{2}  \label{J_w_ud}
\end{equation}

Obviously, we should have $\mathbf{J}_{\pm }\mathbf{=0}$. This inconsistency
arises due to a lack in the modelization relative to the singular case of
infinite wall. Note that, if dealing with a finite barrier, $\left( \gamma
/2\hbar \right) \left\vert \mathbf{A}\right\vert ^{2}\left( q+q^{\prime
}\right) ^{2}=\left( 2\gamma _{c}/\hbar \right) \left\vert \mathbf{A}%
\right\vert ^{2}\mathbf{Q}\left( q+q^{\prime }\right) ^{2}/K^{2}$, where $q$
and $Q$ are small. In the case of an infinite well, we are in a situation
where $K$ tends to infinity. Because of this inconsistency (the infinite
well cannot meet the criteria used in our approximations), this term should
certainly be discarded. The problem can also be circumvented when building
the function%
\begin{equation}
\Phi =\phi \uparrow +\widehat{K}\left( \phi \uparrow \right) =\phi \uparrow
+\phi ^{\ast }\downarrow =2\sin \left( \frac{n\pi }{a}z\right) \left[
iAe^{-i\delta qz}\uparrow +\left( iA\right) ^{\ast }e^{i\delta qz}\downarrow %
\right]  \label{func_plan_w}
\end{equation}%
which properly describes a solution with a spin lying in the plane
perpendicular to the DP field and for which $\mathbf{J}=0$.

\section{$\left[ 100\right] $-oriented barrier, zeroth-order wave-function
coefficients\label{Coeff_0}}

The continuity of the wave function defined by Eqs. \ref{001_100} and of its
derivative at $z=0$ and $z=a$ for the two spin channels provides the
following linear system: 
\begin{subequations}
\label{Perel_100}
\begin{eqnarray}
\hspace{-1cm}-B_{1}+A_{2}+B_{2}-\dfrac{i\xi }{2K^{\prime }}\tilde{A}_{2}+%
\dfrac{i\xi }{2K^{\prime }}\tilde{B}_{2} &=&A_{1}  \label{Perel_100.1} \\
\hspace{-1cm}i\dfrac{q}{K}B_{1}-A_{2}+B_{2}+\dfrac{i\xi }{2K}\tilde{A}_{2}+%
\dfrac{i\xi }{2K}\tilde{B}_{2} &=&i\dfrac{q}{K}A_{1}  \label{Perel_100.2} \\
A_{2}e^{-Ka}+B_{2}e^{Ka}-\dfrac{i\xi }{2K^{\prime }}\tilde{A}%
_{2}e^{-K^{\prime }a}+\dfrac{i\xi }{2K^{\prime }}\tilde{B}_{2}e^{K^{\prime
}a}-A_{3}e^{iqa} &=&0  \label{Perel_100.3} \\
\hspace{-1cm}-A_{2}e^{-Ka}+B_{2}e^{Ka}+\dfrac{i\xi }{2K}\tilde{A}%
_{2}e^{-K^{\prime }a}+\dfrac{i\xi }{2K}\tilde{B}_{2}e^{K^{\prime }a}-i\dfrac{%
q}{K}A_{3}e^{iqa} &=&0  \label{Perel_100.4} \\
\hspace{-1cm}-\tilde{B}_{1}+\dfrac{i\xi }{2K}A_{2}-\dfrac{i\xi }{2K}B_{2}+%
\tilde{A}_{2}+\tilde{B}_{2} &=&\tilde{A}_{1}  \label{Perel_100.5} \\
\hspace{-1cm}i\dfrac{q}{K^{\prime }}\tilde{B}_{1}-\dfrac{i\xi }{2K^{\prime }}%
A_{2}-\dfrac{i\xi }{2K^{\prime }}B_{2}-\tilde{A}_{2}+\tilde{B}_{2} &=&i%
\dfrac{q}{K^{\prime }}\tilde{A}_{1}  \label{Perel_100.6} \\
\hspace{-1cm}\dfrac{i\xi }{2K}A_{2}e^{-Ka}-\dfrac{i\xi }{2K}B_{2}e^{Ka}+%
\tilde{A}_{2}e^{-K^{\prime }a}+\tilde{B}_{2}e^{K^{\prime }a}-\tilde{A}%
_{3}e^{iqa} &=&0  \label{Perel_100.7} \\
\hspace{-1cm}-\dfrac{i\xi }{2K^{\prime }}A_{2}e^{-Ka}-\dfrac{i\xi }{%
2K^{\prime }}B_{2}e^{Ka}-\tilde{A}_{2}e^{-K^{\prime }a}+\tilde{B}%
_{2}e^{K^{\prime }a}-i\dfrac{q}{K^{\prime }}\tilde{A}_{3}e^{iqa} &=&0
\label{Perel_100.8}
\end{eqnarray}

The coefficients $A_{1}$ and $\widetilde{A}_{1}$ which define the intensity
of the two spin components of the incident wave are known (initial
conditions). It could be verified that the determinant of this system is non
zero. We can calculate the eight coefficients $B_{1}$, $\widetilde{B}_{1}$, $%
A_{2}$, $B_{2}$, $\widetilde{A}_{2}$, $\widetilde{B}_{2}$, $A_{3}$, and $%
\widetilde{A}_{3}$ from the eight relations, Eqs. \ref{Perel_100}. We begin
to solve these eight equations to the zeroth order in $\xi /K_{0}$ or, in
other words, by writing $\xi /K_{0}=0$. We note the eight equations are then
divided into two sets: The first four equations are uncoupled to the last
four ones.

The first four equations are related to the spin $\left[ 1\;1\right] ^{t}$
and write as:

\end{subequations}
\begin{subequations}
\label{Perel_110}
\begin{eqnarray}
A_{1} &=&-B_{1}+A_{2}+B_{2}  \label{Perel_110.1} \\
iqA_{1} &=&iqB_{1}-KA_{2}+KB_{2}  \label{Perel_110.2} \\
A_{3}e^{iqa} &=&A_{2}e^{-Ka}+B_{2}e^{Ka}  \label{Perel_110.3} \\
iqA_{3}e^{iqa} &=&-KA_{2}e^{-Ka}+KB_{2}e^{Ka}  \label{Perel_110.4}
\end{eqnarray}%
and the last four ones are related to the spin $\left[ 1\;-1\right] ^{t}$.
The equations are the same by altering $\left( A_{1}\text{, }B_{1}\text{, }%
A_{2}\text{, }B_{2}\text{, }K\right) $ into $\left( \widetilde{A}_{1}\text{, 
}\widetilde{B}_{1}\text{, }\widetilde{A}_{2}\text{, }\widetilde{B}_{2}\text{%
, }K^{\prime }\right) $.

This is the usual formulation of the tunnel effect. Because Eqs. \ref%
{Perel_100} are written to the first order in $\xi /K_{0}$, we are looking
for a solution to the same order.

To give an example, we look for the results when the incident wave has a
spin $\left[ 1\;1\right] ^{t}$ $\left( A_{1}\neq 0\text{, }\widetilde{A}%
_{1}=0\right) $. Considering Eqs. \ref{ab_total}, we note that the
approximation given by the last term of each equation is almost valid as
soon as $Ka>2$. In Ref. 
%TCIMACRO{\TeXButton{\onlinecite{Perel03}}{\onlinecite{Perel03}}}%
%BeginExpansion
\onlinecite{Perel03}%
%EndExpansion
, $K$ is of the order of $0.1$\ \AA $^{-1}$ which gives $a$ of the order of
20 \AA\ in order that the inequality holds, a value which is quite
reasonable.

As\textit{\ }$\widetilde{A}_{1}=0$, this shows that to the zeroth order in $%
\xi /K_{0}$, the results may be summarized by

\end{subequations}
\begin{equation}
\begin{array}{ccc}
A_{2}/A_{1}=f_{2}^{\left( 0\right) } & \qquad \qquad & \widetilde{A}_{2}=0
\\ 
A_{3}/A_{1}=f_{3}^{\left( 0\right) } &  & \widetilde{A}_{3}=0 \\ 
B_{1}/A_{1}=g_{1}^{\left( 0\right) } &  & \widetilde{B}_{1}=0 \\ 
B_{2}/A_{1}=g_{2}^{\left( 0\right) } &  & \widetilde{B}_{2}=0%
\end{array}
\label{Perel_appr0}
\end{equation}%
where $f_{j}^{\left( 0\right) }=f_{j}^{\left( 0\right) }\left( q,K\right) $
and $g_{j}^{\left( 0\right) }=g_{j}^{\left( 0\right) }\left( q,K\right) $
correspond to the standard case (Subsec. \ref{standard}) and can be deduced
from Eqs. \ref{ab_total}. This means that, up to the first order in $\xi
/K_{0}$, the results are of the shape:%
\begin{equation}
\begin{array}{ccc}
A_{2}/A_{1}=f_{2}^{\left( 0\right) }+\left( \xi /K\right) \,f_{2}^{\left(
1\right) } & \qquad \qquad & \widetilde{A}_{2}/A_{1}=\left( \xi /K\right) \,%
\widetilde{f}_{2}^{\left( 1\right) } \\ 
A_{3}/A_{1}=f_{3}^{\left( 0\right) }+\left( \xi /K\right) \,f_{3}^{\left(
1\right) } &  & \widetilde{A}_{3}/A_{1}=\left( \xi /K\right) \,\widetilde{f}%
_{3}^{\left( 1\right) } \\ 
B_{1}/A_{1}=g_{1}^{\left( 0\right) }+\left( \xi /K\right) \,g_{1}^{\left(
1\right) } &  & \widetilde{B}_{1}/A_{1}=\left( \xi /K\right) \,\widetilde{g}%
_{1}^{\left( 1\right) } \\ 
B_{2}/A_{1}=g_{2}^{\left( 0\right) }+\left( \xi /K\right) \,g_{2}^{\left(
1\right) } &  & \widetilde{B}_{2}/A_{1}=\left( \xi /K\right) \,\widetilde{g}%
_{2}^{\left( 1\right) }%
\end{array}
\label{Perel_appr1}
\end{equation}%
where the factors of $f_{j}^{\left( 1\right) }=f_{j}^{\left( 1\right)
}\left( q,K,K^{\prime }\right) $, $g_{j}^{\left( 1\right) }=g_{j}^{\left(
1\right) }\left( q,K,K^{\prime }\right) $, $\widetilde{f}_{j}^{\left(
1\right) }=\widetilde{f}_{j}^{\left( 1\right) }\left( q,K,K^{\prime }\right) 
$, and $\widetilde{g}_{j}^{\left( 1\right) }=\widetilde{g}_{j}^{\left(
1\right) }\left( q,K,K^{\prime }\right) $ may be equal to zero. In fact, a
calculation up to the first order in $\xi /K_{0}$ via Eqs. \ref{Perel_100},
involves terms of $\left( \xi /K\right) \widetilde{A}_{2}$ type, which are
of second-order in $\xi /K_{0}$. Therefore $f_{j}^{\left( 1\right)
}=g_{j}^{\left( 1\right) }=0$ and Eqs. \ref{Perel_appr1} write as

\begin{equation}
\begin{array}{ccc}
A_{2}/A_{1}=f_{2}^{\left( 0\right) }+\left( \xi /K\right)
^{2}\,f_{2}^{\left( 2\right) } & \qquad \qquad & \widetilde{A}%
_{2}/A_{1}=\left( \xi /K\right) \,\widetilde{f}_{2}^{\left( 1\right) } \\ 
A_{3}/A_{1}=f_{3}^{\left( 0\right) }+\left( \xi /K\right)
^{2}\,f_{3}^{\left( 2\right) } &  & \widetilde{A}_{3}/A_{1}=\left( \xi
/K\right) \,\widetilde{f}_{3}^{\left( 1\right) } \\ 
B_{1}/A_{1}=g_{1}^{\left( 0\right) }+\left( \xi /K\right)
^{2}\,g_{1}^{\left( 2\right) } &  & \widetilde{B}_{1}/A_{1}=\left( \xi
/K\right) \,\widetilde{g}_{1}^{\left( 1\right) } \\ 
B_{2}/A_{1}=g_{2}^{\left( 0\right) }+\left( \xi /K\right)
^{2}\,g_{2}^{\left( 2\right) } &  & \widetilde{B}_{2}/A_{1}=\left( \xi
/K\right) \,\widetilde{g}_{2}^{\left( 1\right) }%
\end{array}
\label{Perel_app012}
\end{equation}%
where\textbf{\textit{\ }}$f_{j}^{\left( 2\right) }=f_{j}^{\left( 2\right)
}\left( q,K,K^{\prime }\right) $ and $g_{j}^{\left( 2\right) }=g_{j}^{\left(
2\right) }\left( q,K,K^{\prime }\right) $.

Of course if $A_{1}=0$ and $\widetilde{A}_{1}\neq 0$, the results are to be
inverted. $f_{j}^{(2)}$ $\left( \text{resp. }g_{j}^{(2)}\right) $ is
comparable to, or smaller than, $f_{j}^{(0)}$ $\left( \text{resp. }%
g_{j}^{(0)}\right) $. In Sec. \ref{first}, it is shown that $\widetilde{f}%
_{j}^{\left( 1\right) }A_{1}$ is of the order of $A_{j}$ and $\widetilde{g}%
_{j}^{\left( 1\right) }A_{1}$ is of the order of $B_{j}$.\newpage

\begin{center}
FIGURES
\end{center}

FIG. 1. Sketch of the tunnel geometry with definition of notations. The
spin-orbit-split barrier material of thickness $a$ $($medium $II)$ is
located between two free-electron like materials $($media $I$ and $III)$.
The tunnel axis, normal to the barrier, is the $z$ axis. In the
free-electron-like materials, the real electron wave vector in the $z$
direction is referred to as $\mathbf{q}$. In the barrier material, the
evanescent wave vector along the $z$ axis is referred to as $\mathbf{Q}+i%
\mathbf{K}$, where $\mathbf{Q}$ and $\mathbf{K}$ are real quantities. The
transverse wave vector component, in the barrier plane, $\mathbf{\xi }$ is
conserved in the tunnel process. Then, the overall wave vectors in the three
media are respectively $\mathbf{k}_{I}=\mathbf{k}_{III}=\mathbf{\xi +q}$ and 
$\mathbf{k}_{II}=\mathbf{\xi +Q+}i\mathbf{K}$.

FIG. 2.\textbf{\ }This figure illustrates transformations which, starting
from a state of wave vector $\mathbf{k}$ and with a mean value of the Pauli
operator $\left\langle \widehat{\mathbf{\sigma }}\right\rangle $, construct
degenerate states. $\widehat{K}_{0}$ is the complex conjugation and $%
\widehat{K}=-i\sigma _{y}\widehat{K}_{0}$ is the Kramers time-reversal
operator. $\widehat{K}$ yields a state with the wave vector $-\mathbf{k}%
^{\ast }$ and with the mean spin $-\left\langle \widehat{\mathbf{\sigma }}%
\right\rangle $. The state of wave vector $\mathbf{k}^{\mathbf{\ast }}$ may
be associated to another spin state, corresponding to the mean value $%
\left\langle \widehat{\mathbf{\sigma }}\right\rangle ^{\prime }$. Applying $%
\widehat{K}$ to this state, we form a degenerate state with the wave vector $%
-\mathbf{k}$ and associated to the mean spin $-\left\langle \widehat{\mathbf{%
\sigma }}\right\rangle ^{\prime }$. Four states are finally obtained.

FIG. 3. Plot of the real-energy lines inside the gap for $\mathbf{k}=[\xi
,0,iK]$, where $K$ and $\xi $ are real and positive and $\tan \theta =\xi /K$%
. The calculation is performed using a $14\times 14$ $\mathbf{k\cdot p}$
Hamiltonian. The loops are drawn versus $\left\Vert \mathbf{k}\right\Vert $
in $2\pi /a_{0}$ units, where $a_{0}$ is the cubic lattice parameter. In all
these directions, the spin degeneracy is lifted. Their shape and extension
sharply depend on $\theta $. For $\theta =43.2%
%TCIMACRO{\U{b0}}%
%BeginExpansion
{{}^\circ}%
%EndExpansion
$, the two branches are too close to be resolved at this scale. The
parameters used in the calculation are: $P=9.88$ eV.\AA , $P^{\prime }=0.41$
eV.\AA , $E_{G}=1.519$ eV, $\Delta _{c}=E_{\Gamma 8c}-E_{\Gamma 7c}=0.171$
eV, $P_{X}=8.68$ eV.\AA , $\Delta =0.341$ eV, $E_{\Delta }=E_{\Gamma
7c}-E_{\Gamma 6c}=2.969$ eV, $\Delta ^{\prime }=-0.17$ eV (see Ref. 
%TCIMACRO{\TeXButton{\onlinecite{Jancu}}{\onlinecite{Jancu}} }%
%BeginExpansion
\onlinecite{Jancu}
%EndExpansion
for a complete discussion). Inset: Real-band structure (left, dashed line.
For clarity, only a valence band which is connected to the evanescent branch
is drawn) and evanescent band across the band gap (right, full line) along
the $[001]$ direction $\left( \theta =0\right) $ where the DP
\textquotedblleft exchange\textquotedblright\ field is zero (no spin
splitting).

FIG. 4. Mathematical plot of the real-energy lines for $\mathbf{k}$ along $%
[110]$ as a function of the real part of the wave vector $Q$ in the barrier.
The calculation is performed for a ratio $\gamma /\gamma _{c}=0.438$\ \AA .
We are only concerned with negative energies, which refer to evanescent
states. More precisely, the physical states are located within a very small
energy domain below the origin. The domain $Q>0$ refers to up-spin states,
whereas the domain $Q<0$ refers to down-spin states. In each case, the
imaginary component of the wave vector can take the values $\pm K$. Thus, at
a given energy, we have exactly the \textit{four} possible states $(Q\pm
iK)\uparrow $ and $(-Q\pm iK)\downarrow $. The down-spin states are Kramers
conjugates of the up-spin states.

FIG. 5.\textbf{\ }This figure is a special case of Fig. 2, when the DP field 
$\mathbf{\chi }$ lies along a real direction $\mathbf{n}$. Following the
same procedure, four degenerate states are constructed, which now have their
spins quantized along the \textit{same} direction $\mathbf{n}$ (i.e. $%
\left\langle \widehat{\mathbf{\sigma }}\right\rangle =\left\langle \widehat{%
\mathbf{\sigma }}\right\rangle ^{\prime }$)\textbf{.}

FIG. 6.\textbf{\ (}Color online).\textbf{\ }The lower part of this figure
illustrates the spin-dependent tunneling scheme in the case of a $[001]$
oriented barrier (Perel's case). The horizontal plane describes the electron
wave vector in the barrier; $\mathbf{K}$ is taken along the $[001]$ axis and 
$\mathbf{\xi }$ lies in the barrier plane, along $[100]$. The upper part of
the figure $(E>0)$ corresponds to the real conduction band - the wave
vectors are real quantities - and the parabola-like curves describing
spin-split states along the $[101]$-direction are drawn. An up-spin state
(full line, open circle) with the wave vector $\mathbf{q}^{\prime }$ is
degenerate with a down-spin state at the wave vector $\mathbf{q}$\ (dotted
line, dark circle) and also with up- and down-spin states at the wave
vectors $-\mathbf{q}$ and $-\mathbf{q}^{\prime }$ respectively. This is
useful for the calculation of a quantum well, given in Appendix \ref{IQW110}%
.\ Concerning the evanescent states, in a naive effective-mass picture, one
may think of evanescent states being mirrors of these real states (in the $%
E<0$ domain) with imaginary wave vectors. Then up- and down-spin electrons
at the energy $E$, would tunnel with the two different wave vectors $i%
\mathbf{q}^{\prime }$ and $i\mathbf{q}$, thus resulting in a spin-filter
effect.\ However, our calculation shows that, concerning evanescent states
(lower part of the figure, $E<0$), the situation is not so simple. In the
negative-energy region, the $\mathbf{K}$ axis refers to the imaginary
wave-vector component and $\mathbf{\xi }$ refers to the real wave-vector
component. Real-energy lines are found only when $\tan \theta =\xi /K<1$ .
These real-energy lines, when drawn for a given $\theta $, consist of loops
connecting opposite spin states at the zone center (\textquotedblleft
up\textquotedblright\ spin: Full curve; \textquotedblleft
down\textquotedblright\ spin: Dotted curve. Obviously, when going off the
zone center, the spin no longer remains a good quantum number - in fact, it
can be calculated that its average value rotates along the loop - but it has
to be pointed out that, in the D'yakonov and Perel' description, the energy
eigenvectors are also pure spin states which depend on the $\theta $ ratio).
Two of these loops are drawn here. Let us consider a tunneling process at
the energy $E$ (horizontal grey plane or yellow plane in the online edition)
of an electron with the wave vector component $\mathbf{\xi }$ in the barrier
plane, which has to be conserved in the tunneling process. It can be
observed here that the two states marked on the loops by a dark circle $%
\left( \mathbf{K}^{\prime }\right) $ and an open circle $\left( \mathbf{K}%
\right) $ - which are energy degenerate - are associated to the same real
wave-vector component $\mathbf{\xi }$. However, they correspond to two
different $\theta $ as they are respectively associated to the imaginary
components $i\mathbf{K}$ and $i\mathbf{K}^{\prime }$, along the tunneling
direction. The difference between $K$ and $K^{\prime }$ results in a
spin-filter effect. Inset (upper left): Top view of the plane at energy $E$
showing the intercepts with the loops which determine the relevant wave
vectors $\mathbf{K}$ and $\mathbf{K}^{\prime }$.


\begin{thebibliography}{99}
\bibitem{imag} The imaginary part of $\mathcal{E}\left( \mathbf{k}\right) $
has no physical meaning and should not be confused with the self-energy
related to the lifetime e.g. in many electron calculations.
\bibitem{filter} For a pure spin filter along a given quantization axis, the
transmission operator is a diagonal $2\times 2$ matrix with eigenvalues of
different modulus. Then, we deal with two uncoupled spin channels with a
non-zero transmission asymmetry.
\end{thebibliography}
\end{document}